\documentclass[%
 reprint,
 amsmath,amssymb,
 aps,
]{revtex4-2}

\usepackage{graphicx}
\usepackage{dcolumn}
\usepackage{bm}

\usepackage{siunitx}
\usepackage{physics}
\usepackage{float}

\begin{document}

\preprint{APS/123-QED}

\title{Laser-cavity locking at the $10^{-7}$ instability scale utilizing beam elipticity}

\author{Fritz Diorico}
\author{Artem Zhutov}%
\author{Onur Hosten}%
 \email{onur.hosten@ist.ac.at}
\affiliation{%
Institute of Science and Technology Austria, Klosterneuburg, Austria
}%

\date{\today}

\begin{abstract}
Ultrastable lasers form the back bone of precision measurements in science and technology. Such lasers attain their stability through frequency locking to reference cavities. State-of-the-art locking performances to date had been achieved using frequency-modulation based methods, complemented with active drift cancellation systems. We demonstrate an all passive, modulation-free laser-cavity locking technique (squash locking) that utilizes changes in beam ellipticity for error signal generation, and a coherent polarization post-selection for noise resilience. By comparing two identically built proof-of-principle systems, we show a frequency locking instability of $5 \times 10^{-7}$ relative to the cavity linewidth at \SI{10}{\second} averaging. The results surpass the demonstrated performances of methods engineered over the last five decades, opening a new path for further advancing the precision and simplicity of laser frequency stabilization.
\end{abstract}

\maketitle


Laser frequency stabilization is indispensable in the science and engineering of atomic time keeping \cite{3_Bloom2014}, gravitational wave detection \cite{4_Abbot2016}, tests of relativity \cite{5_Guerlebeck2018}, atom interferometry \cite{6_Rudolph2020}, and in the quantum control of various systems such as atoms \cite{7_Hosten2016}, nanoparticles \cite{8_Kuhn2015} and mechanical oscillators \cite{9_Aspelmeyer2014} – to name a few. As an example, contemporary atomic clocks require milliHertz-linewidth lasers to probe long-lived optical atomic transitions, and the level of precision in stabilization of laser frequencies to reference optical cavities play a crucial role in reaching the state-of-the-art performance levels \cite{10_Kessler2012,11_Matei2017}. A number of methods have been developed over the decades to address the task of locking lasers to cavities. These include: side-of-fringe intensity methods \cite{12_Barger1973}, polarization based methods \cite{13_Hansch1980,14_Asenbaum2011,15_Moriwaki2009}, frequency modulation techniques including transmission modulation \cite{16_White1965,17_Salomon1988} and the Pound-Drever-Hall (PDH) reflection method \cite{1_Drever1983}, and lastly, spatial mode interference methods \cite{18_Wieman1982,19_Shaddock1999,20_Miller2014,21_Zullo2016}.

\begin{figure}[!b]
\includegraphics{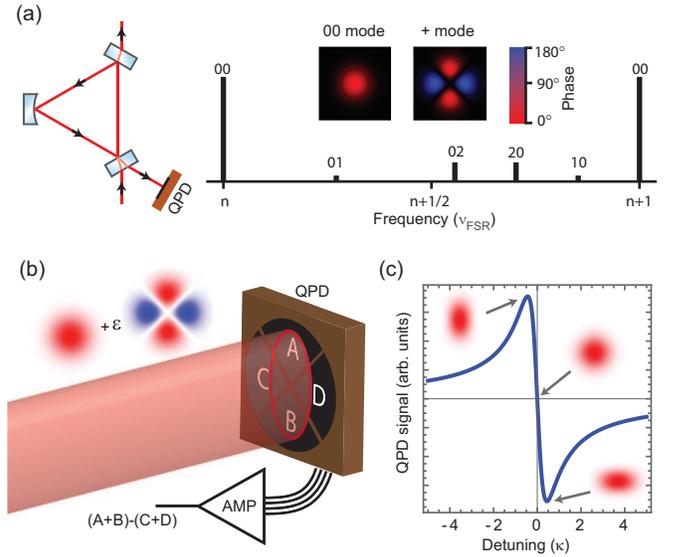}
\caption{\label{fig:illustration} Conceptual illustration. (a) Geometry of the utilized cavity, and its relevant modes (mirrors: two plane, one 5-\SI{}{\centi\meter} radius-of-curvature). Inset: spatial amplitude structure of the ‘00’ and the ‘+’ modes. In terms of the native cavity modes, the ‘+’ mode appears as a superposition of the ‘02’ and ‘20’ modes. Alignment drifts populate the ‘01’ and ‘10’ modes. QPD: quadrant photodiode; $\nu_{_{\mathrm{FSR}}}$: free spectral range; $n$: an integer. (b) Beam ellipticity detection using a QPD, and mode decomposition of a slightly elliptical beam. $\epsilon$ : a small complex number; AMP: transimpedance amplifier. (c) Laser frequency dependence of the QPD signal near the ‘00’ mode resonance. $\kappa$ : cavity full-linewidth.}
\end{figure}

Certain methods could be preferable based on application, but due to its stability and versatility, the PDH technique – utilizing radio-frequency modulation/demodulation of an optical carrier – has become a general standard. Achieving the most demanding locking stabilities of 1 part in $10^5$ - $10^6$ of a cavity linewidth has further required additional layers of active feedback mechanisms to reduce the residual amplitude (RAM) modulation \cite{22_Zhang2014} which typically limits the lock point stability of a PDH setup. A purely passive and reduced-complexity method that could compete with these highly engineered setups could be beneficial for all applications, especially when electro- and acousto-optic devices used for modulation are particularly undesirable \cite{23_Sheard2010,24_Swierad2016}.

Here we develop a precision laser frequency stabilization scheme that utilizes completely passive optical elements. The main method consists of monitoring the change in the ellipticity of a beam reflected from a cavity. This method is further enhanced with a pre- and a post-selection of the beam polarization to coherently suppress technical noise that limits system performance. Notably, through this enhancement technique, we also uncover a curious physical phenomenon where the post-selection gives rise to an effective cavity with variable loss or gain.

Our scheme differs from previous spatial mode methods in its robustness of implementation and insensitivity to alignment drifts. Further, it differs from previous polarization based methods in its utilization of polarization only to reject noise. These aspects enable us to realize the true potential of modulation-free schemes. Both aspects of the scheme are cavity geometry independent, applicable to linear or ring cavities (see Supplemental Material (SM) \cite{Supp}), and can be used independently of each other.

To understand the main stabilization method, one needs to recall the Hermite-Gaussian (HG) spatial modes supported by an optical cavity \cite{25_Siegman1986}. Here, of interest are the fundamental HG modes labeled ‘00’, and specific second-order HG modes that we label ‘+’ (Fig. \ref{fig:illustration}a). A slightly elliptical beam with a horizontal/vertical orientation can mathematically be decomposed into a main ‘00’ component and a small ‘+’ component (Fig. \ref{fig:illustration}b). For such a beam, the phase difference between these two components encodes the information about ellipticity. When incident on the cavity near a ‘00’ resonance, only the ‘00’ component builds up inside the cavity, acquiring a phase shift in the process, and giving rise to a change in the relative phase between the two components outside of the cavity. Through this mode-dependent phase shift, the reflected beam can be made to acquire opposite ellipticities on opposite sides of the resonance (Fig. \ref{fig:illustration}c). From the perspective of cavity design, the only requirement for proper operation is for the ‘00’ mode not to be spectrally degenerate with the second-order modes. To harness this effect we use a quadrant photodiode (QPD), subtracting the sums of the diagonals to obtain an error signal proportional to beam ellipticity (Fig. \ref{fig:illustration}b\&c; see SM \cite{Supp}). Specifically, the generated error signal is given by the imaginary part of the reflection coefficient in Eqn. \ref{eq:1}.

\begin{figure}[b]
\includegraphics{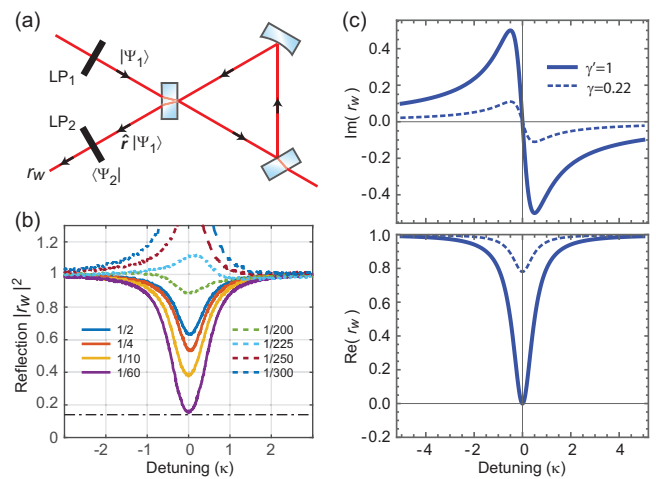}
\caption{\label{fig:weakvalue} Weak value enhancement: Effective cavity under polarization pre- and post-selection. (a) Input/output polarizers are used to obtain the weak-value $r_w$ of the reflection operator  $\hat{r}$, resulting in an effective cavity with tunable reflection. LP: Linear polarizer. (b) Measured normalized reflection power for different post-selection configurations. Curves are labeled by their off-resonant remaining power fraction  $|\braket{\psi_2}{\psi_1}|^2$. $1/2$-curve: same as original cavity reflection curve; $\sim 1/60$-curve: effective impedance matched ($\gamma'=1$) configuration. Dash-dot line: Intensity contribution level from the non-‘00’ mode components. (c) Real and imaginary parts of $r_w$  for an input mode matched to the ‘00’ mode (theory), illustrating the advantages of post-selection: Effective increase in the slope of the error signal, and minimization of optical power on the QPD near the locking point.}
\end{figure}

\begin{figure}[b]
\includegraphics{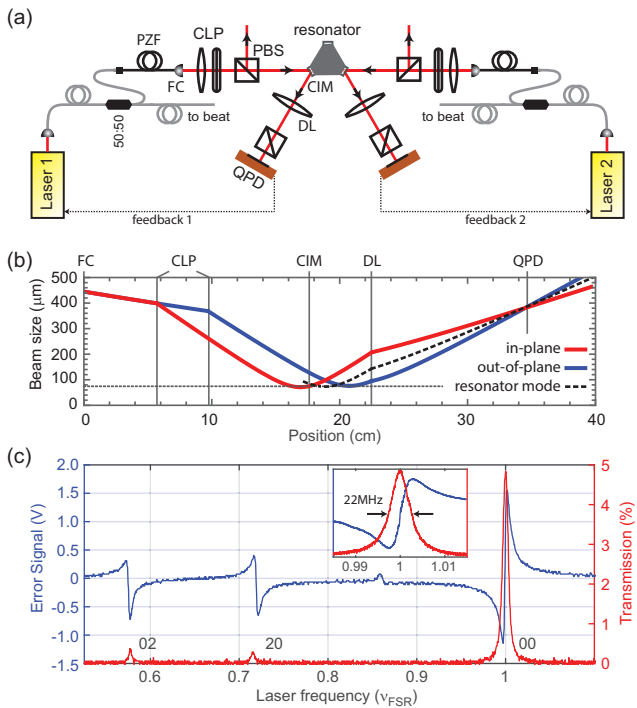}
\caption{\label{fig:experimentalsetup} Experimental setup. (a) Setup for locking two identical lasers to a common optical cavity. At the beatnote setup (not shown), one path is frequency shifted by \SI{80}{\mega\hertz}   with an acousto-optic modulator before combining the paths on a beam splitter leading to a fast photodiode. Except the lasers, the setup is enclosed in a metal box for temperature stability and reduced air flow. CLP: cylindrical lens pair, 150-\SI{}{\milli\meter} convex each, one oriented with axis in-plane the other out-of-plane; FC: fiber collimator with adjustable focus; ‘50:50’: polarization maintaining fiber splitter; PZF: polarizing fiber; PBS: rotatable polarizing beam splitter; CIM: cavity input mirror; DL: detection lens, 100-\SI{}{\milli\meter} convex spherical; QPD: quadrant photodiode (Fig. \ref{fig:illustration}b). (b) Location of the relevant elements, and resulting beam properties. Black dotted line: 74-\SI{}{\micro\meter} mean cavity mode waist (\SI{84}{\micro\meter} in-plane, \SI{65}{\micro\meter} um out-of-plane). Actual DL position is immaterial, DL to QPD distance matters. (c) Typical error signal along with cavity transmission.}
\end{figure}

\begin{figure*}
\includegraphics{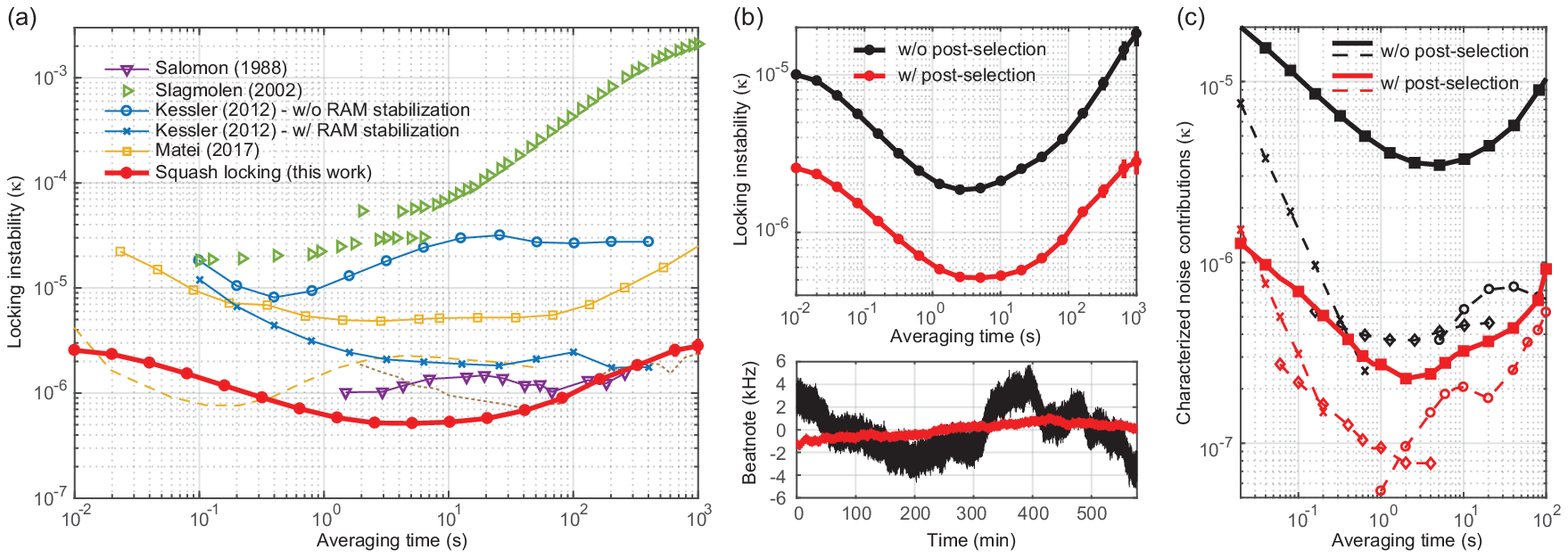}
\caption{\label{fig:lockperformance}Laser frequency locking performance. (a) Achieved frequency locking instability for an individual laser relative to cavity linewidth (filled circles). Allan deviation of a 10-hour recorded beatnote is taken and equal instabilities for the two locks is assumed. Comparison of locking instability with previous work: Salomon \cite{17_Salomon1988}: transmission modulation method; Slagmolen \cite{38_Slagmolen2002}: double-passed ‘tilt locking’ method; Matei \cite{11_Matei2017}: PDH method; Kessler \cite{10_Kessler2012}: PDH method out-of-loop noise floor characterization with and without RAM stabilization (arXiv:1112.3854 for this data). Data are extracted and expressed relative to cavity linewidhts utilized in each work (see SM \cite{Supp}). Dashed and dotted lines: see text. (b) Top: Comparison of achieved instability for operation with (red) and without (black) post-selection, bottom: corresponding beatnote traces. Throughout, error bars indicate $\pm 1$ standard deviation. (c) Estimated limitations to stability. Red: operation with post-selection, black: without post-selection; solid lines: dominant contributions due to beam shape noise, dashed lines: subdominant contributions. Break-down of subdominant contributions: measured in-loop error signal noise (crosses), QPD dark noise (diamonds) and laser intensity drift noise (circles).  }
\end{figure*}

The detected signal can be thought of as an interference between the resonating ‘00’ mode that leaks out of the cavity, and the second-order ‘+’ mode that promptly reflects from the cavity to form a phase reference. This detection modality is insensitive to alignment drifts and fluctuations, since small misalignments simply generate first-order modes in the mode decomposition \cite{19_Shaddock1999}. A limitation to stability is still posed, for example, by residual fluctuations in the incident beam shape, attributable to the alteration of second-order mode components – affecting the phase reference. To help alleviate this residual limitation, we utilize the polarization pre- and post-selection procedure, which we efficiently describe using the weak-value concept \cite{2_Aharonov1988}. This concept has been utilized over the last decade to amplify small signals in optical settings \cite{26_Hosten2008,27_Dixon2009,28_Starling2010,29_Xu2013} with mostly a good understanding of its benefits \cite{30_Jordan2014,31_Torres2016}.

Intuitively, the post-selection process allows one to directly perform a differential measurement between the signals that would be generated by the two polarization components of light incident on the cavity – through interference of amplitudes from the two polarizations. With polarization-resolved cavity resonances, one polarization can build up resonantly in the cavity while the other polarization reflects promptly, serving as a reference for the noise originating prior to the cavity. The differential measurement implemented by the post-selection then isolates the cavity signal. The post-selection setting adapted below enforces a complete destructive interference between the two amplitudes when the cavity is exactly on resonance. The description of this phenomenon takes the form of an effective loss-tunable cavity.

With the polarization degree of freedom included, the cavity reflection coefficient is replaced by a reflection operator $\hat{r}$  acting on the input polarization state $\ket{\psi_1}$. When a post-selection onto state  $\ket{\psi_2}$ is made, the resulting effective reflection coefficient  $r_w$   (Fig. \ref{fig:weakvalue}a) is given by the weak-value of the reflection operator (see SM \cite{Supp}):

\begin{equation}
\label{eq:1}
r_w = \frac{\bra{\psi_2}\hat{r}\ket{\psi_1}}{\bra{\psi_2}\ket{\psi_1}} = 1 - \gamma' \frac{1}{1-i\frac{\delta}{\kappa/2}}
\end{equation}
Here $\delta$  is the laser-cavity frequency detuning, $\kappa$  is the full-width cavity linewidth, and $\gamma'=A\gamma$. The parameter $\gamma$ is a dimensionless constant characterizing the roundtrip losses in the cavity, where $\gamma <1$, $\gamma=1$, $\gamma>1$ and $\gamma>2$ can be identified respectively with the under-coupled, impedance-matched, over-coupled, and cavity-gain regimes. The ‘amplification’ parameter $A$ depends on the specific post-selection, and it is real-valued for linear polarization states $\ket{\psi_1}$  and  $\ket{\psi_2}$. While it is a function of polarizer angles, operationally it can be related to the remaining power fraction  $|\bra{\psi_1}\ket{\psi_2}| \approx 1 / (2A+1)^2$ after post-selection when the light is off-resonant (see SM \cite{Supp}). The right hand side of Eqn. \ref{eq:1} is of the same form as the reflection from a regular high-finesse cavity, except with a variable loss parameter $\gamma'$  (at fixed  $\kappa$) instead if the fixed parameter $\gamma$. We demonstrate the variability of $\gamma'$ experimentally in Figure \ref{fig:weakvalue}b by observing the post-selection dependence of the reflected power from the cavity. 

Tuning the post-selection to the effective impedance-matched (Fig. \ref{fig:weakvalue}c) configuration reduces noise for frequency stabilization purposes. In this configuration, the non-signal-generating beam components incident on the QPD are strongly attenuated while signal generating parts are attenuated less, effectively increasing error signal slope and reducing sensitivity to beam shape or intensity fluctuations (see SM \cite{Supp}).

To demonstrate the performance of our stabilization scheme, we build two identical copies of the locking setup (Fig. \ref{fig:experimentalsetup}a), and stabilize two separate lasers to two degenerate counter-propagating ‘00’ modes of a triangular ring cavity (see SM \cite{Supp}). To assess stability, we split-off 50\% of the power from the lasers to monitor the beating frequency of the lasers on a fast photodiode using a frequency counter (SRS FS740).

The implementation of the locking scheme relies on beam shaping for achieving opposite beam ellipticities on opposite sides of the resonance (Fig. \ref{fig:illustration}c). This goal requires the phase difference between the ‘00’ and the ‘+’ components of the beam at the QPD location to be $\pi/2$ radians when the frequency is matched to the ‘00’ resonance. In this configuration the interference of the two components results into a circular beam profile at the QPD, yielding a zero error signal. When the frequency is detuned from resonance, the reflected ‘00’ component acquires an additional phase shift determined by Eqn. \ref{eq:1}, directly altering the phase difference at the QPD location. For detunings of opposite sign, the phase-shifted interference results into opposite beam ellipticities, yielding opposite sign error signals. Note that the phase shift recedes to its on-resonance value for large detunings, rendering the beam circular once more far off-resonance. This allows us to do alignment and tuning while off resonance.

The required beam shaping is achieved using a cylindrical lens pair (CLP), through which a circular input beam is astigmatically focused. The distance between the CLP determines the amount of light in the ‘+’ mode, amounting to ~10\% of the total power (\SI{400}{\micro\watt}) in our demonstration. Lens positions are aligned such that, as the beam propagates it reaches back to a circular profile at the QPD location when the light is off-resonance (Fig. \ref{fig:experimentalsetup}b). The size of this circular beam needs to match that of the forward-propagated cavity ‘00’ mode to maximize the signal. For additional details on the design principle of the CLP and its relation to the error signal size, see SM \cite{Supp}.

The cavity utilized in this work has a full-linewidth of 22-\SI{}{\mega\hertz} and a finesse of $195$ for the ‘00’ mode (and a corresponding free spectral range of $\nu_{_{\mathrm{FSR}}} = \SI{4.28}{\giga\hertz}$). Two independent external cavity diode lasers at \SI{780}{\nano\meter} are employed to probe the cavity. Each laser has an intrinsic-linewidth of \SI{500}{\kilo\hertz}. For a discussion on the effect of laser linewidths on performance, see SM \cite{Supp}. A typical error signal from this setup is shown in Figure \ref{fig:experimentalsetup}c as one varies the frequency of one of the lasers. For reference, the largest peak of the signal here corresponds to a beam aspect ratio of about 1.3. To independently lock the frequencies of the two lasers to a common resonance, the obtained error signals are fed back to the currents of the respective lasers through feedback controllers (see SM \cite{Supp}). 

A comparison between different optical frequency stabilization systems can be made by characterizing the performance as a fractional instability with respect to the cavity linewidth – nominally, absolute instabilities will scale proportional to the cavity linewidth since typical sources of problematic instabilities originate from fluctuations in the error signal shape. In Figure \ref{fig:lockperformance}a, we present the performance of the implemented scheme while locked to a ‘00’ mode resonance, and compare it to the best known implementations of other stabilization methods. It can be seen that we enter the $10^{-7}$ locking instability regime for averaging times between \SI{0.25}{\second} and \SI{100}{\second}, surpassing previous state-of-the-art, and reaching $5 \times 10^{-7}$ instability at \SI{10}{\second} averaging. The explicit improvement brought by the post-selection is also illustrated in Figure \ref{fig:lockperformance}b. Note that prior to this work, the tightest demonstrated laser-cavity locking was achieved using the transmission modulation method \cite{17_Salomon1988} (Fig. \ref{fig:lockperformance}a). For the PDH method, although recent engineering of RAM levels that are almost compatible with our obtained instabilities were reported \cite{32_Matei2016,33_Jin2021} (Fig. \ref{fig:lockperformance}a, dashed and dotted lines respectively), a demonstration of an actual laser-cavity locking at such levels had not been reported.

A characterization of various noise sources (Fig. \ref{fig:lockperformance}c; see SM \cite{Supp}) reveals that the residual fluctuations of the incident beam shape constitutes the dominant limitation to the currently achieved instability level. A major source of this instability is the thermal-stress induced polarization changes inside the launching fibers, which we observed to directly translate into beam ellipticity fluctuations. In fact, we note that switching from polarization maintaining fibers to polarizing fibers (Fig. \ref{fig:experimentalsetup}a) had provided nearly an order of magnitude improvement, allowing us to reach the current results. Further stability improvements will require better control over the input beam shape stability. For the effects of the long-term temperature drifts on stability, see SM \cite{Supp}.

A generic optical-cavity application can integrate ‘squash locking’ as a plug-and-play method, benefiting from simplicity, robustness and performance. The lack of RF modulation, which avoids spectral contamination and electromagnetic interference, makes the technique particularly suited for applications like optical frequency conversion \cite{34_Shaddock2000} or stabilization of laser injection locking \cite{35_Ottaway2001}. For high-performance applications, such as space-based laser ranging (including gravitational wave detection) or development of optical frequency standards \cite{11_Matei2017}, there are often additional requirements which the technique is capable of delivering: Direct compatibility with low-light level measurements (see SM \cite{Supp}), and direct integration compatibility with ultralow-vibration-sensitivity cavities (see SM \cite{Supp}). Recently, utilizing the ‘tilt locking’  technique \cite{19_Shaddock1999}, a modulation-free laser locking system for ranging applications was demonstrated \cite{36_Chabbra2021}, showing compatibility with demanding space mission requirements. With the current technique, such systems could be implemented or improved more easily. Lastly, given its direct compatibility with on-chip systems such as dielectric whispering-gallery cavities (see SM \cite{Supp}), the technique can benefit industrial optical communication systems requiring narrow linewidth lasers \cite{37_AlTaiy2014}.

\begin{acknowledgments}
This work is supported by IST Austria. We thank Rishabh Sahu and Sebastian Wald for technical contributions to the experiment.
\end{acknowledgments}


\nocite{*}

\bibliography{squash}

\providecommand{\noopsort}[1]{}\providecommand{\singleletter}[1]{#1}%
\begin{thebibliography}{39}%
\makeatletter
\providecommand \@ifxundefined [1]{%
 \@ifx{#1\undefined}
}%
\providecommand \@ifnum [1]{%
 \ifnum #1\expandafter \@firstoftwo
 \else \expandafter \@secondoftwo
 \fi
}%
\providecommand \@ifx [1]{%
 \ifx #1\expandafter \@firstoftwo
 \else \expandafter \@secondoftwo
 \fi
}%
\providecommand \natexlab [1]{#1}%
\providecommand \enquote  [1]{``#1''}%
\providecommand \bibnamefont  [1]{#1}%
\providecommand \bibfnamefont [1]{#1}%
\providecommand \citenamefont [1]{#1}%
\providecommand \href@noop [0]{\@secondoftwo}%
\providecommand \href [0]{\begingroup \@sanitize@url \@href}%
\providecommand \@href[1]{\@@startlink{#1}\@@href}%
\providecommand \@@href[1]{\endgroup#1\@@endlink}%
\providecommand \@sanitize@url [0]{\catcode `\\12\catcode `\$12\catcode
  `\&12\catcode `\#12\catcode `\^12\catcode `\_12\catcode `\%12\relax}%
\providecommand \@@startlink[1]{}%
\providecommand \@@endlink[0]{}%
\providecommand \url  [0]{\begingroup\@sanitize@url \@url }%
\providecommand \@url [1]{\endgroup\@href {#1}{\urlprefix }}%
\providecommand \urlprefix  [0]{URL }%
\providecommand \Eprint [0]{\href }%
\providecommand \doibase [0]{https://doi.org/}%
\providecommand \selectlanguage [0]{\@gobble}%
\providecommand \bibinfo  [0]{\@secondoftwo}%
\providecommand \bibfield  [0]{\@secondoftwo}%
\providecommand \translation [1]{[#1]}%
\providecommand \BibitemOpen [0]{}%
\providecommand \bibitemStop [0]{}%
\providecommand \bibitemNoStop [0]{.\EOS\space}%
\providecommand \EOS [0]{\spacefactor3000\relax}%
\providecommand \BibitemShut  [1]{\csname bibitem#1\endcsname}%
\let\auto@bib@innerbib\@empty
\bibitem [{\citenamefont {Bloom}\ \emph {et~al.}(2014)\citenamefont {Bloom},
  \citenamefont {Nicholson}, \citenamefont {Williams}, \citenamefont
  {Campbell}, \citenamefont {Bishof}, \citenamefont {Zhang}, \citenamefont
  {Zhang}, \citenamefont {Bromley},\ and\ \citenamefont {Ye}}]{3_Bloom2014}%
  \BibitemOpen
  \bibfield  {author} {\bibinfo {author} {\bibfnamefont {B.~J.}\ \bibnamefont
  {Bloom}}, \bibinfo {author} {\bibfnamefont {T.~L.}\ \bibnamefont
  {Nicholson}}, \bibinfo {author} {\bibfnamefont {J.~R.}\ \bibnamefont
  {Williams}}, \bibinfo {author} {\bibfnamefont {S.~L.}\ \bibnamefont
  {Campbell}}, \bibinfo {author} {\bibfnamefont {M.}~\bibnamefont {Bishof}},
  \bibinfo {author} {\bibfnamefont {X.}~\bibnamefont {Zhang}}, \bibinfo
  {author} {\bibfnamefont {W.}~\bibnamefont {Zhang}}, \bibinfo {author}
  {\bibfnamefont {S.~L.}\ \bibnamefont {Bromley}},\ and\ \bibinfo {author}
  {\bibfnamefont {J.}~\bibnamefont {Ye}},\ }\bibfield  {title} {\bibinfo
  {title} {An optical lattice clock with accuracy and stability at the
  10\textsuperscript{-18} level},\ }\href {https://doi.org/10.1038/nature12941}
  {\bibfield  {journal} {\bibinfo  {journal} {Nature}\ }\textbf {\bibinfo
  {volume} {506}},\ \bibinfo {pages} {71} (\bibinfo {year} {2014})}\BibitemShut
  {NoStop}%
\bibitem [{\citenamefont {Abbott}\ \emph {et~al.}(2016)\citenamefont {Abbott}
  \emph {et~al.}}]{4_Abbot2016}%
  \BibitemOpen
  \bibfield  {author} {\bibinfo {author} {\bibfnamefont {B.}~\bibnamefont
  {Abbott}} \emph {et~al.} (\bibinfo {collaboration} {LIGO Scientific
  Collaboration and Virgo Collaboration}),\ }\bibfield  {title} {\bibinfo
  {title} {Observation of gravitational waves from a binary black hole
  merger},\ }\href {https://doi.org/10.1103/PhysRevLett.116.061102} {\bibfield
  {journal} {\bibinfo  {journal} {Phys. Rev. Lett.}\ }\textbf {\bibinfo
  {volume} {116}},\ \bibinfo {pages} {061102} (\bibinfo {year}
  {2016})}\BibitemShut {NoStop}%
\bibitem [{\citenamefont {G\"urlebeck}\ \emph {et~al.}(2018)\citenamefont
  {G\"urlebeck}, \citenamefont {W\"orner}, \citenamefont {Schuldt},
  \citenamefont {D\"oringshoff}, \citenamefont {Gaul}, \citenamefont {Gerardi},
  \citenamefont {Grenzebach}, \citenamefont {Jha}, \citenamefont {Kovalchuk},
  \citenamefont {Resch}, \citenamefont {Wendrich}, \citenamefont {Berger},
  \citenamefont {Herrmann}, \citenamefont {Johann}, \citenamefont {Krutzik},
  \citenamefont {Peters}, \citenamefont {Rasel},\ and\ \citenamefont
  {Braxmaier}}]{5_Guerlebeck2018}%
  \BibitemOpen
  \bibfield  {author} {\bibinfo {author} {\bibfnamefont {N.}~\bibnamefont
  {G\"urlebeck}}, \bibinfo {author} {\bibfnamefont {L.}~\bibnamefont
  {W\"orner}}, \bibinfo {author} {\bibfnamefont {T.}~\bibnamefont {Schuldt}},
  \bibinfo {author} {\bibfnamefont {K.}~\bibnamefont {D\"oringshoff}}, \bibinfo
  {author} {\bibfnamefont {K.}~\bibnamefont {Gaul}}, \bibinfo {author}
  {\bibfnamefont {D.}~\bibnamefont {Gerardi}}, \bibinfo {author} {\bibfnamefont
  {A.}~\bibnamefont {Grenzebach}}, \bibinfo {author} {\bibfnamefont
  {N.}~\bibnamefont {Jha}}, \bibinfo {author} {\bibfnamefont {E.}~\bibnamefont
  {Kovalchuk}}, \bibinfo {author} {\bibfnamefont {A.}~\bibnamefont {Resch}},
  \bibinfo {author} {\bibfnamefont {T.}~\bibnamefont {Wendrich}}, \bibinfo
  {author} {\bibfnamefont {R.}~\bibnamefont {Berger}}, \bibinfo {author}
  {\bibfnamefont {S.}~\bibnamefont {Herrmann}}, \bibinfo {author}
  {\bibfnamefont {U.}~\bibnamefont {Johann}}, \bibinfo {author} {\bibfnamefont
  {M.}~\bibnamefont {Krutzik}}, \bibinfo {author} {\bibfnamefont
  {A.}~\bibnamefont {Peters}}, \bibinfo {author} {\bibfnamefont {E.~M.}\
  \bibnamefont {Rasel}},\ and\ \bibinfo {author} {\bibfnamefont
  {C.}~\bibnamefont {Braxmaier}},\ }\bibfield  {title} {\bibinfo {title}
  {Boost: A satellite mission to test lorentz invariance using high-performance
  optical frequency references},\ }\href
  {https://doi.org/10.1103/PhysRevD.97.124051} {\bibfield  {journal} {\bibinfo
  {journal} {Phys. Rev. D}\ }\textbf {\bibinfo {volume} {97}},\ \bibinfo
  {pages} {124051} (\bibinfo {year} {2018})}\BibitemShut {NoStop}%
\bibitem [{\citenamefont {Rudolph}\ \emph {et~al.}(2020)\citenamefont
  {Rudolph}, \citenamefont {Wilkason}, \citenamefont {Nantel}, \citenamefont
  {Swan}, \citenamefont {Holland}, \citenamefont {Jiang}, \citenamefont
  {Garber}, \citenamefont {Carman},\ and\ \citenamefont
  {Hogan}}]{6_Rudolph2020}%
  \BibitemOpen
  \bibfield  {author} {\bibinfo {author} {\bibfnamefont {J.}~\bibnamefont
  {Rudolph}}, \bibinfo {author} {\bibfnamefont {T.}~\bibnamefont {Wilkason}},
  \bibinfo {author} {\bibfnamefont {M.}~\bibnamefont {Nantel}}, \bibinfo
  {author} {\bibfnamefont {H.}~\bibnamefont {Swan}}, \bibinfo {author}
  {\bibfnamefont {C.~M.}\ \bibnamefont {Holland}}, \bibinfo {author}
  {\bibfnamefont {Y.}~\bibnamefont {Jiang}}, \bibinfo {author} {\bibfnamefont
  {B.~E.}\ \bibnamefont {Garber}}, \bibinfo {author} {\bibfnamefont {S.~P.}\
  \bibnamefont {Carman}},\ and\ \bibinfo {author} {\bibfnamefont {J.~M.}\
  \bibnamefont {Hogan}},\ }\bibfield  {title} {\bibinfo {title} {Large momentum
  transfer clock atom interferometry on the 689 nm intercombination line of
  strontium},\ }\href {https://doi.org/10.1103/PhysRevLett.124.083604}
  {\bibfield  {journal} {\bibinfo  {journal} {Phys. Rev. Lett.}\ }\textbf
  {\bibinfo {volume} {124}},\ \bibinfo {pages} {083604} (\bibinfo {year}
  {2020})}\BibitemShut {NoStop}%
\bibitem [{\citenamefont {Hosten}\ \emph {et~al.}(2016)\citenamefont {Hosten},
  \citenamefont {Engelsen}, \citenamefont {Krishnakumar},\ and\ \citenamefont
  {Kasevich}}]{7_Hosten2016}%
  \BibitemOpen
  \bibfield  {author} {\bibinfo {author} {\bibfnamefont {O.}~\bibnamefont
  {Hosten}}, \bibinfo {author} {\bibfnamefont {N.~J.}\ \bibnamefont
  {Engelsen}}, \bibinfo {author} {\bibfnamefont {R.}~\bibnamefont
  {Krishnakumar}},\ and\ \bibinfo {author} {\bibfnamefont {M.~A.}\ \bibnamefont
  {Kasevich}},\ }\bibfield  {title} {\bibinfo {title} {Measurement noise 100
  times lower than the quantum-projection limit using entangled atoms},\ }\href
  {https://doi.org/10.1038/nature16176} {\bibfield  {journal} {\bibinfo
  {journal} {Nature}\ }\textbf {\bibinfo {volume} {529}},\ \bibinfo {pages}
  {505} (\bibinfo {year} {2016})}\BibitemShut {NoStop}%
\bibitem [{\citenamefont {Kuhn}\ \emph {et~al.}(2015)\citenamefont {Kuhn},
  \citenamefont {Asenbaum}, \citenamefont {Kosloff}, \citenamefont {Sclafani},
  \citenamefont {Stickler}, \citenamefont {Nimmrichter}, \citenamefont
  {Hornberger}, \citenamefont {Cheshnovsky}, \citenamefont {Patolsky},\ and\
  \citenamefont {Arndt}}]{8_Kuhn2015}%
  \BibitemOpen
  \bibfield  {author} {\bibinfo {author} {\bibfnamefont {S.}~\bibnamefont
  {Kuhn}}, \bibinfo {author} {\bibfnamefont {P.}~\bibnamefont {Asenbaum}},
  \bibinfo {author} {\bibfnamefont {A.}~\bibnamefont {Kosloff}}, \bibinfo
  {author} {\bibfnamefont {M.}~\bibnamefont {Sclafani}}, \bibinfo {author}
  {\bibfnamefont {B.~A.}\ \bibnamefont {Stickler}}, \bibinfo {author}
  {\bibfnamefont {S.}~\bibnamefont {Nimmrichter}}, \bibinfo {author}
  {\bibfnamefont {K.}~\bibnamefont {Hornberger}}, \bibinfo {author}
  {\bibfnamefont {O.}~\bibnamefont {Cheshnovsky}}, \bibinfo {author}
  {\bibfnamefont {F.}~\bibnamefont {Patolsky}},\ and\ \bibinfo {author}
  {\bibfnamefont {M.}~\bibnamefont {Arndt}},\ }\bibfield  {title} {\bibinfo
  {title} {Cavity-assisted manipulation of freely rotating silicon nanorods in
  high vacuum},\ }\href@noop {} {\bibfield  {journal} {\bibinfo  {journal}
  {Nano Letters}\ }\textbf {\bibinfo {volume} {15}},\ \bibinfo {pages} {5604}
  (\bibinfo {year} {2015})}\BibitemShut {NoStop}%
\bibitem [{\citenamefont {Aspelmeyer}\ \emph {et~al.}(2014)\citenamefont
  {Aspelmeyer}, \citenamefont {Kippenberg},\ and\ \citenamefont
  {Marquardt}}]{9_Aspelmeyer2014}%
  \BibitemOpen
  \bibfield  {author} {\bibinfo {author} {\bibfnamefont {M.}~\bibnamefont
  {Aspelmeyer}}, \bibinfo {author} {\bibfnamefont {T.~J.}\ \bibnamefont
  {Kippenberg}},\ and\ \bibinfo {author} {\bibfnamefont {F.}~\bibnamefont
  {Marquardt}},\ }\bibfield  {title} {\bibinfo {title} {Cavity optomechanics},\
  }\href {https://doi.org/10.1103/RevModPhys.86.1391} {\bibfield  {journal}
  {\bibinfo  {journal} {Rev. Mod. Phys.}\ }\textbf {\bibinfo {volume} {86}},\
  \bibinfo {pages} {1391} (\bibinfo {year} {2014})}\BibitemShut {NoStop}%
\bibitem [{\citenamefont {Kessler}\ \emph {et~al.}(2012)\citenamefont
  {Kessler}, \citenamefont {Hagemann}, \citenamefont {Grebing}, \citenamefont
  {Legero}, \citenamefont {Sterr}, \citenamefont {Riehle}, \citenamefont
  {Martin}, \citenamefont {Chen},\ and\ \citenamefont {Ye}}]{10_Kessler2012}%
  \BibitemOpen
  \bibfield  {author} {\bibinfo {author} {\bibfnamefont {T.}~\bibnamefont
  {Kessler}}, \bibinfo {author} {\bibfnamefont {C.}~\bibnamefont {Hagemann}},
  \bibinfo {author} {\bibfnamefont {C.}~\bibnamefont {Grebing}}, \bibinfo
  {author} {\bibfnamefont {T.}~\bibnamefont {Legero}}, \bibinfo {author}
  {\bibfnamefont {U.}~\bibnamefont {Sterr}}, \bibinfo {author} {\bibfnamefont
  {F.}~\bibnamefont {Riehle}}, \bibinfo {author} {\bibfnamefont {M.~J.}\
  \bibnamefont {Martin}}, \bibinfo {author} {\bibfnamefont {L.}~\bibnamefont
  {Chen}},\ and\ \bibinfo {author} {\bibfnamefont {J.}~\bibnamefont {Ye}},\
  }\bibfield  {title} {\bibinfo {title} {A sub-40-m{H}z-linewidth laser based
  on a silicon single-crystal optical cavity},\ }\href
  {https://doi.org/10.1038/nphoton.2012.217} {\bibfield  {journal} {\bibinfo
  {journal} {Nature Photonics}\ }\textbf {\bibinfo {volume} {6}},\ \bibinfo
  {pages} {687} (\bibinfo {year} {2012})}\BibitemShut {NoStop}%
\bibitem [{\citenamefont {Matei}\ \emph {et~al.}(2017)\citenamefont {Matei},
  \citenamefont {Legero}, \citenamefont {H\"afner}, \citenamefont {Grebing},
  \citenamefont {Weyrich}, \citenamefont {Zhang}, \citenamefont {Sonderhouse},
  \citenamefont {Robinson}, \citenamefont {Ye}, \citenamefont {Riehle},\ and\
  \citenamefont {Sterr}}]{11_Matei2017}%
  \BibitemOpen
  \bibfield  {author} {\bibinfo {author} {\bibfnamefont {D.~G.}\ \bibnamefont
  {Matei}}, \bibinfo {author} {\bibfnamefont {T.}~\bibnamefont {Legero}},
  \bibinfo {author} {\bibfnamefont {S.}~\bibnamefont {H\"afner}}, \bibinfo
  {author} {\bibfnamefont {C.}~\bibnamefont {Grebing}}, \bibinfo {author}
  {\bibfnamefont {R.}~\bibnamefont {Weyrich}}, \bibinfo {author} {\bibfnamefont
  {W.}~\bibnamefont {Zhang}}, \bibinfo {author} {\bibfnamefont
  {L.}~\bibnamefont {Sonderhouse}}, \bibinfo {author} {\bibfnamefont {J.~M.}\
  \bibnamefont {Robinson}}, \bibinfo {author} {\bibfnamefont {J.}~\bibnamefont
  {Ye}}, \bibinfo {author} {\bibfnamefont {F.}~\bibnamefont {Riehle}},\ and\
  \bibinfo {author} {\bibfnamefont {U.}~\bibnamefont {Sterr}},\ }\bibfield
  {title} {\bibinfo {title} {$1.5\text{ }\ensuremath{\mu}\mathrm{m}$ lasers
  with sub-10 m{H}z linewidth},\ }\href
  {https://doi.org/10.1103/PhysRevLett.118.263202} {\bibfield  {journal}
  {\bibinfo  {journal} {Phys. Rev. Lett.}\ }\textbf {\bibinfo {volume} {118}},\
  \bibinfo {pages} {263202} (\bibinfo {year} {2017})}\BibitemShut {NoStop}%
\bibitem [{\citenamefont {Barger}\ \emph {et~al.}(1973)\citenamefont {Barger},
  \citenamefont {Sorem},\ and\ \citenamefont {Hall}}]{12_Barger1973}%
  \BibitemOpen
  \bibfield  {author} {\bibinfo {author} {\bibfnamefont {R.~L.}\ \bibnamefont
  {Barger}}, \bibinfo {author} {\bibfnamefont {M.}~\bibnamefont {Sorem}},\ and\
  \bibinfo {author} {\bibfnamefont {J.}~\bibnamefont {Hall}},\ }\bibfield
  {title} {\bibinfo {title} {Frequency stabilization of a cw dye laser},\
  }\href {https://doi.org/10.1063/1.1654513} {\bibfield  {journal} {\bibinfo
  {journal} {Applied Physics Letters}\ }\textbf {\bibinfo {volume} {22}},\
  \bibinfo {pages} {573} (\bibinfo {year} {1973})}\BibitemShut {NoStop}%
\bibitem [{\citenamefont {H\"ansch}\ and\ \citenamefont
  {Couillaud}(1980)}]{13_Hansch1980}%
  \BibitemOpen
  \bibfield  {author} {\bibinfo {author} {\bibfnamefont {T.}~\bibnamefont
  {H\"ansch}}\ and\ \bibinfo {author} {\bibfnamefont {B.}~\bibnamefont
  {Couillaud}},\ }\bibfield  {title} {\bibinfo {title} {Laser frequency
  stabilization by polarization spectroscopy of a reflecting reference
  cavity},\ }\href
  {https://doi.org/https://doi.org/10.1016/0030-4018(80)90069-3} {\bibfield
  {journal} {\bibinfo  {journal} {Optics Communications}\ }\textbf {\bibinfo
  {volume} {35}},\ \bibinfo {pages} {441} (\bibinfo {year} {1980})}\BibitemShut
  {NoStop}%
\bibitem [{\citenamefont {Asenbaum}\ and\ \citenamefont
  {Arndt}(2011)}]{14_Asenbaum2011}%
  \BibitemOpen
  \bibfield  {author} {\bibinfo {author} {\bibfnamefont {P.}~\bibnamefont
  {Asenbaum}}\ and\ \bibinfo {author} {\bibfnamefont {M.}~\bibnamefont
  {Arndt}},\ }\bibfield  {title} {\bibinfo {title} {Cavity stabilization using
  the weak intrinsic birefringence of dielectric mirrors},\ }\href
  {https://doi.org/10.1364/OL.36.003720} {\bibfield  {journal} {\bibinfo
  {journal} {Opt. Lett.}\ }\textbf {\bibinfo {volume} {36}},\ \bibinfo {pages}
  {3720} (\bibinfo {year} {2011})}\BibitemShut {NoStop}%
\bibitem [{\citenamefont {{Moriwaki}}\ \emph {et~al.}(2009)\citenamefont
  {{Moriwaki}}, \citenamefont {{Mori}}, \citenamefont {{Takeno}},\ and\
  \citenamefont {{Mio}}}]{15_Moriwaki2009}%
  \BibitemOpen
  \bibfield  {author} {\bibinfo {author} {\bibfnamefont {S.}~\bibnamefont
  {{Moriwaki}}}, \bibinfo {author} {\bibfnamefont {T.}~\bibnamefont {{Mori}}},
  \bibinfo {author} {\bibfnamefont {K.}~\bibnamefont {{Takeno}}},\ and\
  \bibinfo {author} {\bibfnamefont {N.}~\bibnamefont {{Mio}}},\ }\bibfield
  {title} {\bibinfo {title} {{Frequency Discrimination Method Making Use of
  Polarization Selectivity of Triangular Optical Cavity}},\ }\href
  {https://doi.org/10.1143/APEX.2.016501} {\bibfield  {journal} {\bibinfo
  {journal} {Applied Physics Express}\ }\textbf {\bibinfo {volume} {2}},\
  \bibinfo {eid} {016501} (\bibinfo {year} {2009})}\BibitemShut {NoStop}%
\bibitem [{\citenamefont {White}(1965)}]{16_White1965}%
  \BibitemOpen
  \bibfield  {author} {\bibinfo {author} {\bibfnamefont {A.}~\bibnamefont
  {White}},\ }\bibfield  {title} {\bibinfo {title} {Frequency stabilization of
  gas lasers},\ }\href {https://doi.org/10.1109/JQE.1965.1072246} {\bibfield
  {journal} {\bibinfo  {journal} {IEEE Journal of Quantum Electronics}\
  }\textbf {\bibinfo {volume} {1}},\ \bibinfo {pages} {349} (\bibinfo {year}
  {1965})}\BibitemShut {NoStop}%
\bibitem [{\citenamefont {Salomon}\ \emph {et~al.}(1988)\citenamefont
  {Salomon}, \citenamefont {Hils},\ and\ \citenamefont
  {Hall}}]{17_Salomon1988}%
  \BibitemOpen
  \bibfield  {author} {\bibinfo {author} {\bibfnamefont {C.}~\bibnamefont
  {Salomon}}, \bibinfo {author} {\bibfnamefont {D.}~\bibnamefont {Hils}},\ and\
  \bibinfo {author} {\bibfnamefont {J.~L.}\ \bibnamefont {Hall}},\ }\bibfield
  {title} {\bibinfo {title} {Laser stabilization at the millihertz level},\
  }\href {https://doi.org/10.1364/JOSAB.5.001576} {\bibfield  {journal}
  {\bibinfo  {journal} {J. Opt. Soc. Am. B}\ }\textbf {\bibinfo {volume} {5}},\
  \bibinfo {pages} {1576} (\bibinfo {year} {1988})}\BibitemShut {NoStop}%
\bibitem [{\citenamefont {Drever}\ \emph {et~al.}(1983)\citenamefont {Drever},
  \citenamefont {Hall}, \citenamefont {Kowalski}, \citenamefont {Hough},
  \citenamefont {Ford}, \citenamefont {Munley},\ and\ \citenamefont
  {Ward}}]{1_Drever1983}%
  \BibitemOpen
  \bibfield  {author} {\bibinfo {author} {\bibfnamefont {R.~W.~P.}\
  \bibnamefont {Drever}}, \bibinfo {author} {\bibfnamefont {J.~L.}\
  \bibnamefont {Hall}}, \bibinfo {author} {\bibfnamefont {F.~V.}\ \bibnamefont
  {Kowalski}}, \bibinfo {author} {\bibfnamefont {J.}~\bibnamefont {Hough}},
  \bibinfo {author} {\bibfnamefont {G.~M.}\ \bibnamefont {Ford}}, \bibinfo
  {author} {\bibfnamefont {A.~J.}\ \bibnamefont {Munley}},\ and\ \bibinfo
  {author} {\bibfnamefont {H.}~\bibnamefont {Ward}},\ }\bibfield  {title}
  {\bibinfo {title} {Laser phase and frequency stabilization using an optical
  resonator},\ }\href {https://doi.org/10.1007/BF00702605} {\bibfield
  {journal} {\bibinfo  {journal} {Applied Physics B}\ }\textbf {\bibinfo
  {volume} {31}},\ \bibinfo {pages} {97} (\bibinfo {year} {1983})}\BibitemShut
  {NoStop}%
\bibitem [{\citenamefont {Wieman}\ and\ \citenamefont
  {Gilbert}(1982)}]{18_Wieman1982}%
  \BibitemOpen
  \bibfield  {author} {\bibinfo {author} {\bibfnamefont {C.~E.}\ \bibnamefont
  {Wieman}}\ and\ \bibinfo {author} {\bibfnamefont {S.~L.}\ \bibnamefont
  {Gilbert}},\ }\bibfield  {title} {\bibinfo {title} {Laser-frequency
  stabilization using mode interference from a reflecting reference
  interferometer},\ }\href {https://doi.org/10.1364/OL.7.000480} {\bibfield
  {journal} {\bibinfo  {journal} {Opt. Lett.}\ }\textbf {\bibinfo {volume}
  {7}},\ \bibinfo {pages} {480} (\bibinfo {year} {1982})}\BibitemShut {NoStop}%
\bibitem [{\citenamefont {Shaddock}\ \emph {et~al.}(1999)\citenamefont
  {Shaddock}, \citenamefont {Gray},\ and\ \citenamefont
  {McClelland}}]{19_Shaddock1999}%
  \BibitemOpen
  \bibfield  {author} {\bibinfo {author} {\bibfnamefont {D.~A.}\ \bibnamefont
  {Shaddock}}, \bibinfo {author} {\bibfnamefont {M.~B.}\ \bibnamefont {Gray}},\
  and\ \bibinfo {author} {\bibfnamefont {D.~E.}\ \bibnamefont {McClelland}},\
  }\bibfield  {title} {\bibinfo {title} {Frequency locking a laser to an
  optical cavity by use of spatial mode interference},\ }\href
  {https://doi.org/10.1364/OL.24.001499} {\bibfield  {journal} {\bibinfo
  {journal} {Opt. Lett.}\ }\textbf {\bibinfo {volume} {24}},\ \bibinfo {pages}
  {1499} (\bibinfo {year} {1999})}\BibitemShut {NoStop}%
\bibitem [{\citenamefont {Miller}\ and\ \citenamefont
  {Evans}(2014)}]{20_Miller2014}%
  \BibitemOpen
  \bibfield  {author} {\bibinfo {author} {\bibfnamefont {J.}~\bibnamefont
  {Miller}}\ and\ \bibinfo {author} {\bibfnamefont {M.}~\bibnamefont {Evans}},\
  }\bibfield  {title} {\bibinfo {title} {Length control of an optical resonator
  using second-order transverse modes},\ }\href
  {https://doi.org/10.1364/OL.39.002495} {\bibfield  {journal} {\bibinfo
  {journal} {Opt. Lett.}\ }\textbf {\bibinfo {volume} {39}},\ \bibinfo {pages}
  {2495} (\bibinfo {year} {2014})}\BibitemShut {NoStop}%
\bibitem [{\citenamefont {Zullo}\ \emph {et~al.}(2016)\citenamefont {Zullo},
  \citenamefont {Giorgini}, \citenamefont {Avino}, \citenamefont {Malara},
  \citenamefont {Natale},\ and\ \citenamefont {Gagliardi}}]{21_Zullo2016}%
  \BibitemOpen
  \bibfield  {author} {\bibinfo {author} {\bibfnamefont {R.}~\bibnamefont
  {Zullo}}, \bibinfo {author} {\bibfnamefont {A.}~\bibnamefont {Giorgini}},
  \bibinfo {author} {\bibfnamefont {S.}~\bibnamefont {Avino}}, \bibinfo
  {author} {\bibfnamefont {P.}~\bibnamefont {Malara}}, \bibinfo {author}
  {\bibfnamefont {P.~D.}\ \bibnamefont {Natale}},\ and\ \bibinfo {author}
  {\bibfnamefont {G.}~\bibnamefont {Gagliardi}},\ }\bibfield  {title} {\bibinfo
  {title} {Laser-frequency locking to a whispering-gallery-mode cavity by
  spatial interference of scattered light},\ }\href
  {https://doi.org/10.1364/OL.41.000650} {\bibfield  {journal} {\bibinfo
  {journal} {Opt. Lett.}\ }\textbf {\bibinfo {volume} {41}},\ \bibinfo {pages}
  {650} (\bibinfo {year} {2016})}\BibitemShut {NoStop}%
\bibitem [{\citenamefont {Zhang}\ \emph {et~al.}(2014)\citenamefont {Zhang},
  \citenamefont {Martin}, \citenamefont {Benko}, \citenamefont {Hall},
  \citenamefont {Ye}, \citenamefont {Hagemann}, \citenamefont {Legero},
  \citenamefont {Sterr}, \citenamefont {Riehle}, \citenamefont {Cole},\ and\
  \citenamefont {Aspelmeyer}}]{22_Zhang2014}%
  \BibitemOpen
  \bibfield  {author} {\bibinfo {author} {\bibfnamefont {W.}~\bibnamefont
  {Zhang}}, \bibinfo {author} {\bibfnamefont {M.~J.}\ \bibnamefont {Martin}},
  \bibinfo {author} {\bibfnamefont {C.}~\bibnamefont {Benko}}, \bibinfo
  {author} {\bibfnamefont {J.~L.}\ \bibnamefont {Hall}}, \bibinfo {author}
  {\bibfnamefont {J.}~\bibnamefont {Ye}}, \bibinfo {author} {\bibfnamefont
  {C.}~\bibnamefont {Hagemann}}, \bibinfo {author} {\bibfnamefont
  {T.}~\bibnamefont {Legero}}, \bibinfo {author} {\bibfnamefont
  {U.}~\bibnamefont {Sterr}}, \bibinfo {author} {\bibfnamefont
  {F.}~\bibnamefont {Riehle}}, \bibinfo {author} {\bibfnamefont {G.~D.}\
  \bibnamefont {Cole}},\ and\ \bibinfo {author} {\bibfnamefont
  {M.}~\bibnamefont {Aspelmeyer}},\ }\bibfield  {title} {\bibinfo {title}
  {Reduction of residual amplitude modulation to 1\texttimes
  10\textsuperscript{-6} for frequency modulation and laser stabilization},\
  }\href {https://doi.org/10.1364/OL.39.001980} {\bibfield  {journal} {\bibinfo
   {journal} {Opt. Lett.}\ }\textbf {\bibinfo {volume} {39}},\ \bibinfo {pages}
  {1980} (\bibinfo {year} {2014})}\BibitemShut {NoStop}%
\bibitem [{\citenamefont {Sheard}\ \emph {et~al.}(2010)\citenamefont {Sheard},
  \citenamefont {Heinzel},\ and\ \citenamefont {Danzmann}}]{23_Sheard2010}%
  \BibitemOpen
  \bibfield  {author} {\bibinfo {author} {\bibfnamefont {B.}~\bibnamefont
  {Sheard}}, \bibinfo {author} {\bibfnamefont {G.}~\bibnamefont {Heinzel}},\
  and\ \bibinfo {author} {\bibfnamefont {K.}~\bibnamefont {Danzmann}},\
  }\bibfield  {title} {\bibinfo {title} {{LISA} long-arm interferometry: an
  alternative frequency pre-stabilization system},\ }\href
  {https://doi.org/10.1088/0264-9381/27/8/084011} {\bibfield  {journal}
  {\bibinfo  {journal} {Classical and Quantum Gravity}\ }\textbf {\bibinfo
  {volume} {27}},\ \bibinfo {pages} {084011} (\bibinfo {year}
  {2010})}\BibitemShut {NoStop}%
\bibitem [{\citenamefont {{\'{S}}wierad}\ \emph {et~al.}(2016)\citenamefont
  {{\'{S}}wierad}, \citenamefont {H{\"a}fner}, \citenamefont {Vogt},
  \citenamefont {Venon}, \citenamefont {Holleville}, \citenamefont {Bize},
  \citenamefont {Kulosa}, \citenamefont {Bode}, \citenamefont {Singh},
  \citenamefont {Bongs}, \citenamefont {Rasel}, \citenamefont {Lodewyck},
  \citenamefont {Le~Targat}, \citenamefont {Lisdat},\ and\ \citenamefont
  {Sterr}}]{24_Swierad2016}%
  \BibitemOpen
  \bibfield  {author} {\bibinfo {author} {\bibfnamefont {D.}~\bibnamefont
  {{\'{S}}wierad}}, \bibinfo {author} {\bibfnamefont {S.}~\bibnamefont
  {H{\"a}fner}}, \bibinfo {author} {\bibfnamefont {S.}~\bibnamefont {Vogt}},
  \bibinfo {author} {\bibfnamefont {B.}~\bibnamefont {Venon}}, \bibinfo
  {author} {\bibfnamefont {D.}~\bibnamefont {Holleville}}, \bibinfo {author}
  {\bibfnamefont {S.}~\bibnamefont {Bize}}, \bibinfo {author} {\bibfnamefont
  {A.}~\bibnamefont {Kulosa}}, \bibinfo {author} {\bibfnamefont
  {S.}~\bibnamefont {Bode}}, \bibinfo {author} {\bibfnamefont {Y.}~\bibnamefont
  {Singh}}, \bibinfo {author} {\bibfnamefont {K.}~\bibnamefont {Bongs}},
  \bibinfo {author} {\bibfnamefont {E.~M.}\ \bibnamefont {Rasel}}, \bibinfo
  {author} {\bibfnamefont {J.}~\bibnamefont {Lodewyck}}, \bibinfo {author}
  {\bibfnamefont {R.}~\bibnamefont {Le~Targat}}, \bibinfo {author}
  {\bibfnamefont {C.}~\bibnamefont {Lisdat}},\ and\ \bibinfo {author}
  {\bibfnamefont {U.}~\bibnamefont {Sterr}},\ }\bibfield  {title} {\bibinfo
  {title} {Ultra-stable clock laser system development towards space
  applications},\ }\href {https://doi.org/10.1038/srep33973} {\bibfield
  {journal} {\bibinfo  {journal} {Scientific Reports}\ }\textbf {\bibinfo
  {volume} {6}},\ \bibinfo {pages} {33973} (\bibinfo {year}
  {2016})}\BibitemShut {NoStop}%
\bibitem [{Sup()}]{Supp}%
  \BibitemOpen
  \href@noop {} {\bibinfo  {journal} {See Supplemental Material at XXX}\
  }\BibitemShut {NoStop}%
\bibitem [{\citenamefont {Siegman}(1986)}]{25_Siegman1986}%
  \BibitemOpen
\bibfield  {journal} {  }\bibfield  {author} {\bibinfo {author} {\bibfnamefont
  {A.~E.}\ \bibnamefont {Siegman}},\ }\href@noop {} {\emph {\bibinfo {title}
  {Lasers}}}\ (\bibinfo  {publisher} {University Science Books},\ \bibinfo
  {year} {1986})\BibitemShut {NoStop}%
\bibitem [{\citenamefont {Slagmolen}\ \emph {et~al.}(2002)\citenamefont
  {Slagmolen}, \citenamefont {Shaddock}, \citenamefont {Gray},\ and\
  \citenamefont {McClelland}}]{38_Slagmolen2002}%
  \BibitemOpen
  \bibfield  {author} {\bibinfo {author} {\bibfnamefont {B.}~\bibnamefont
  {Slagmolen}}, \bibinfo {author} {\bibfnamefont {D.}~\bibnamefont {Shaddock}},
  \bibinfo {author} {\bibfnamefont {M.}~\bibnamefont {Gray}},\ and\ \bibinfo
  {author} {\bibfnamefont {D.}~\bibnamefont {McClelland}},\ }\bibfield  {title}
  {\bibinfo {title} {Frequency stability of spatial mode interference (tilt)
  locking},\ }\href {https://doi.org/10.1109/JQE.2002.804267} {\bibfield
  {journal} {\bibinfo  {journal} {IEEE Journal of Quantum Electronics}\
  }\textbf {\bibinfo {volume} {38}},\ \bibinfo {pages} {1521} (\bibinfo {year}
  {2002})}\BibitemShut {NoStop}%
\bibitem [{\citenamefont {Aharonov}\ \emph {et~al.}(1988)\citenamefont
  {Aharonov}, \citenamefont {Albert},\ and\ \citenamefont
  {Vaidman}}]{2_Aharonov1988}%
  \BibitemOpen
  \bibfield  {author} {\bibinfo {author} {\bibfnamefont {Y.}~\bibnamefont
  {Aharonov}}, \bibinfo {author} {\bibfnamefont {D.~Z.}\ \bibnamefont
  {Albert}},\ and\ \bibinfo {author} {\bibfnamefont {L.}~\bibnamefont
  {Vaidman}},\ }\bibfield  {title} {\bibinfo {title} {How the result of a
  measurement of a component of the spin of a spin-1/2 particle can turn out to
  be 100},\ }\href {https://doi.org/10.1103/PhysRevLett.60.1351} {\bibfield
  {journal} {\bibinfo  {journal} {Phys. Rev. Lett.}\ }\textbf {\bibinfo
  {volume} {60}},\ \bibinfo {pages} {1351} (\bibinfo {year}
  {1988})}\BibitemShut {NoStop}%
\bibitem [{\citenamefont {Hosten}\ and\ \citenamefont
  {Kwiat}(2008)}]{26_Hosten2008}%
  \BibitemOpen
  \bibfield  {author} {\bibinfo {author} {\bibfnamefont {O.}~\bibnamefont
  {Hosten}}\ and\ \bibinfo {author} {\bibfnamefont {P.}~\bibnamefont {Kwiat}},\
  }\bibfield  {title} {\bibinfo {title} {{Observation of the Spin Hall Effect
  of Light via Weak Measurements}},\ }\href
  {https://doi.org/10.1126/science.1152697} {\bibfield  {journal} {\bibinfo
  {journal} {Science}\ }\textbf {\bibinfo {volume} {319}},\ \bibinfo {pages}
  {787} (\bibinfo {year} {2008})}\BibitemShut {NoStop}%
\bibitem [{\citenamefont {Dixon}\ \emph {et~al.}(2009)\citenamefont {Dixon},
  \citenamefont {Starling}, \citenamefont {Jordan},\ and\ \citenamefont
  {Howell}}]{27_Dixon2009}%
  \BibitemOpen
  \bibfield  {author} {\bibinfo {author} {\bibfnamefont {P.~B.}\ \bibnamefont
  {Dixon}}, \bibinfo {author} {\bibfnamefont {D.~J.}\ \bibnamefont {Starling}},
  \bibinfo {author} {\bibfnamefont {A.~N.}\ \bibnamefont {Jordan}},\ and\
  \bibinfo {author} {\bibfnamefont {J.~C.}\ \bibnamefont {Howell}},\ }\bibfield
   {title} {\bibinfo {title} {Ultrasensitive beam deflection measurement via
  interferometric weak value amplification},\ }\href
  {https://doi.org/10.1103/PhysRevLett.102.173601} {\bibfield  {journal}
  {\bibinfo  {journal} {Phys. Rev. Lett.}\ }\textbf {\bibinfo {volume} {102}},\
  \bibinfo {pages} {173601} (\bibinfo {year} {2009})}\BibitemShut {NoStop}%
\bibitem [{\citenamefont {Starling}\ \emph {et~al.}(2010)\citenamefont
  {Starling}, \citenamefont {Dixon}, \citenamefont {Jordan},\ and\
  \citenamefont {Howell}}]{28_Starling2010}%
  \BibitemOpen
  \bibfield  {author} {\bibinfo {author} {\bibfnamefont {D.~J.}\ \bibnamefont
  {Starling}}, \bibinfo {author} {\bibfnamefont {P.~B.}\ \bibnamefont {Dixon}},
  \bibinfo {author} {\bibfnamefont {A.~N.}\ \bibnamefont {Jordan}},\ and\
  \bibinfo {author} {\bibfnamefont {J.~C.}\ \bibnamefont {Howell}},\ }\bibfield
   {title} {\bibinfo {title} {Precision frequency measurements with
  interferometric weak values},\ }\href
  {https://doi.org/10.1103/PhysRevA.82.063822} {\bibfield  {journal} {\bibinfo
  {journal} {Phys. Rev. A}\ }\textbf {\bibinfo {volume} {82}},\ \bibinfo
  {pages} {063822} (\bibinfo {year} {2010})}\BibitemShut {NoStop}%
\bibitem [{\citenamefont {Xu}\ \emph {et~al.}(2013)\citenamefont {Xu},
  \citenamefont {Kedem}, \citenamefont {Sun}, \citenamefont {Vaidman},
  \citenamefont {Li},\ and\ \citenamefont {Guo}}]{29_Xu2013}%
  \BibitemOpen
  \bibfield  {author} {\bibinfo {author} {\bibfnamefont {X.-Y.}\ \bibnamefont
  {Xu}}, \bibinfo {author} {\bibfnamefont {Y.}~\bibnamefont {Kedem}}, \bibinfo
  {author} {\bibfnamefont {K.}~\bibnamefont {Sun}}, \bibinfo {author}
  {\bibfnamefont {L.}~\bibnamefont {Vaidman}}, \bibinfo {author} {\bibfnamefont
  {C.-F.}\ \bibnamefont {Li}},\ and\ \bibinfo {author} {\bibfnamefont {G.-C.}\
  \bibnamefont {Guo}},\ }\bibfield  {title} {\bibinfo {title} {Phase estimation
  with weak measurement using a white light source},\ }\href
  {https://doi.org/10.1103/PhysRevLett.111.033604} {\bibfield  {journal}
  {\bibinfo  {journal} {Phys. Rev. Lett.}\ }\textbf {\bibinfo {volume} {111}},\
  \bibinfo {pages} {033604} (\bibinfo {year} {2013})}\BibitemShut {NoStop}%
\bibitem [{\citenamefont {Jordan}\ \emph {et~al.}(2014)\citenamefont {Jordan},
  \citenamefont {Mart\'{\i}nez-Rinc\'on},\ and\ \citenamefont
  {Howell}}]{30_Jordan2014}%
  \BibitemOpen
  \bibfield  {author} {\bibinfo {author} {\bibfnamefont {A.~N.}\ \bibnamefont
  {Jordan}}, \bibinfo {author} {\bibfnamefont {J.}~\bibnamefont
  {Mart\'{\i}nez-Rinc\'on}},\ and\ \bibinfo {author} {\bibfnamefont {J.~C.}\
  \bibnamefont {Howell}},\ }\bibfield  {title} {\bibinfo {title} {Technical
  advantages for weak-value amplification: When less is more},\ }\href
  {https://doi.org/10.1103/PhysRevX.4.011031} {\bibfield  {journal} {\bibinfo
  {journal} {Phys. Rev. X}\ }\textbf {\bibinfo {volume} {4}},\ \bibinfo {pages}
  {011031} (\bibinfo {year} {2014})}\BibitemShut {NoStop}%
\bibitem [{\citenamefont {Torres}\ and\ \citenamefont
  {Salazar-Serrano}(2016)}]{31_Torres2016}%
  \BibitemOpen
  \bibfield  {author} {\bibinfo {author} {\bibfnamefont {J.~P.}\ \bibnamefont
  {Torres}}\ and\ \bibinfo {author} {\bibfnamefont {L.~J.}\ \bibnamefont
  {Salazar-Serrano}},\ }\bibfield  {title} {\bibinfo {title} {Weak value
  amplification: a view from quantum estimation theory that highlights what it
  is and what isn't},\ }\href {https://doi.org/10.1038/srep19702} {\bibfield
  {journal} {\bibinfo  {journal} {Scientific Reports}\ }\textbf {\bibinfo
  {volume} {6}},\ \bibinfo {pages} {19702} (\bibinfo {year}
  {2016})}\BibitemShut {NoStop}%
\bibitem [{\citenamefont {Matei}\ \emph {et~al.}(2016)\citenamefont {Matei},
  \citenamefont {Legero}, \citenamefont {Grebing}, \citenamefont {Häfner},
  \citenamefont {Lisdat}, \citenamefont {Weyrich}, \citenamefont {Zhang},
  \citenamefont {Sonderhouse}, \citenamefont {Robinson}, \citenamefont
  {Riehle}, \citenamefont {Ye},\ and\ \citenamefont {Sterr}}]{32_Matei2016}%
  \BibitemOpen
  \bibfield  {author} {\bibinfo {author} {\bibfnamefont {D.~G.}\ \bibnamefont
  {Matei}}, \bibinfo {author} {\bibfnamefont {T.}~\bibnamefont {Legero}},
  \bibinfo {author} {\bibfnamefont {C.}~\bibnamefont {Grebing}}, \bibinfo
  {author} {\bibfnamefont {S.}~\bibnamefont {Häfner}}, \bibinfo {author}
  {\bibfnamefont {C.}~\bibnamefont {Lisdat}}, \bibinfo {author} {\bibfnamefont
  {R.}~\bibnamefont {Weyrich}}, \bibinfo {author} {\bibfnamefont
  {W.}~\bibnamefont {Zhang}}, \bibinfo {author} {\bibfnamefont
  {L.}~\bibnamefont {Sonderhouse}}, \bibinfo {author} {\bibfnamefont {J.~M.}\
  \bibnamefont {Robinson}}, \bibinfo {author} {\bibfnamefont {F.}~\bibnamefont
  {Riehle}}, \bibinfo {author} {\bibfnamefont {J.}~\bibnamefont {Ye}},\ and\
  \bibinfo {author} {\bibfnamefont {U.}~\bibnamefont {Sterr}},\ }\bibfield
  {title} {\bibinfo {title} {A second generation of low thermal noise cryogenic
  silicon resonators},\ }\href {https://doi.org/10.1088/1742-6596/723/1/012031}
  {\bibfield  {journal} {\bibinfo  {journal} {Journal of Physics: Conference
  Series}\ }\textbf {\bibinfo {volume} {723}},\ \bibinfo {pages} {012031}
  (\bibinfo {year} {2016})}\BibitemShut {NoStop}%
\bibitem [{\citenamefont {Jin}(2021)}]{33_Jin2021}%
  \BibitemOpen
  \bibfield  {author} {\bibinfo {author} {\bibfnamefont {L.}~\bibnamefont
  {Jin}},\ }\bibfield  {title} {\bibinfo {title} {Suppression of residual
  amplitude modulation of adp electro-optical modulator in
  {P}ound-{D}rever-{H}all laser frequency stabilization},\ }\href
  {https://doi.org/https://doi.org/10.1016/j.optlastec.2020.106758} {\bibfield
  {journal} {\bibinfo  {journal} {Optics \& Laser Technology}\ }\textbf
  {\bibinfo {volume} {136}},\ \bibinfo {pages} {106758} (\bibinfo {year}
  {2021})}\BibitemShut {NoStop}%
\bibitem [{\citenamefont {Shaddock}\ \emph {et~al.}(2000)\citenamefont
  {Shaddock}, \citenamefont {Buchler}, \citenamefont {Bowen}, \citenamefont
  {Gray},\ and\ \citenamefont {Lam}}]{34_Shaddock2000}%
  \BibitemOpen
  \bibfield  {author} {\bibinfo {author} {\bibfnamefont {D.~A.}\ \bibnamefont
  {Shaddock}}, \bibinfo {author} {\bibfnamefont {B.~C.}\ \bibnamefont
  {Buchler}}, \bibinfo {author} {\bibfnamefont {W.~P.}\ \bibnamefont {Bowen}},
  \bibinfo {author} {\bibfnamefont {M.~B.}\ \bibnamefont {Gray}},\ and\
  \bibinfo {author} {\bibfnamefont {P.~K.}\ \bibnamefont {Lam}},\ }\bibfield
  {title} {\bibinfo {title} {Modulation-free control of a continuous-wave
  second-harmonic generator},\ }\href
  {https://doi.org/10.1088/1464-4258/2/5/309} {\bibfield  {journal} {\bibinfo
  {journal} {Journal of Optics A: Pure and Applied Optics}\ }\textbf {\bibinfo
  {volume} {2}},\ \bibinfo {pages} {400} (\bibinfo {year} {2000})}\BibitemShut
  {NoStop}%
\bibitem [{\citenamefont {Ottaway}\ \emph {et~al.}(2001)\citenamefont
  {Ottaway}, \citenamefont {Gray}, \citenamefont {Shaddock}, \citenamefont
  {Hollitt}, \citenamefont {Veitch}, \citenamefont {Munch},\ and\ \citenamefont
  {McClelland}}]{35_Ottaway2001}%
  \BibitemOpen
  \bibfield  {author} {\bibinfo {author} {\bibfnamefont {D.}~\bibnamefont
  {Ottaway}}, \bibinfo {author} {\bibfnamefont {M.}~\bibnamefont {Gray}},
  \bibinfo {author} {\bibfnamefont {D.}~\bibnamefont {Shaddock}}, \bibinfo
  {author} {\bibfnamefont {C.}~\bibnamefont {Hollitt}}, \bibinfo {author}
  {\bibfnamefont {P.}~\bibnamefont {Veitch}}, \bibinfo {author} {\bibfnamefont
  {J.}~\bibnamefont {Munch}},\ and\ \bibinfo {author} {\bibfnamefont
  {D.}~\bibnamefont {McClelland}},\ }\bibfield  {title} {\bibinfo {title}
  {Stabilization of injection-locked lasers using spatial mode interference},\
  }\href {https://doi.org/10.1109/3.918577} {\bibfield  {journal} {\bibinfo
  {journal} {IEEE Journal of Quantum Electronics}\ }\textbf {\bibinfo {volume}
  {37}},\ \bibinfo {pages} {653} (\bibinfo {year} {2001})}\BibitemShut
  {NoStop}%
\bibitem [{\citenamefont {Chabbra}\ \emph {et~al.}(2021)\citenamefont
  {Chabbra}, \citenamefont {Wade}, \citenamefont {Rees}, \citenamefont
  {Sutton}, \citenamefont {Stochino}, \citenamefont {Ward}, \citenamefont
  {Shaddock},\ and\ \citenamefont {McKenzie}}]{36_Chabbra2021}%
  \BibitemOpen
  \bibfield  {author} {\bibinfo {author} {\bibfnamefont {N.}~\bibnamefont
  {Chabbra}}, \bibinfo {author} {\bibfnamefont {A.~R.}\ \bibnamefont {Wade}},
  \bibinfo {author} {\bibfnamefont {E.~R.}\ \bibnamefont {Rees}}, \bibinfo
  {author} {\bibfnamefont {A.~J.}\ \bibnamefont {Sutton}}, \bibinfo {author}
  {\bibfnamefont {A.}~\bibnamefont {Stochino}}, \bibinfo {author}
  {\bibfnamefont {R.~L.}\ \bibnamefont {Ward}}, \bibinfo {author}
  {\bibfnamefont {D.~A.}\ \bibnamefont {Shaddock}},\ and\ \bibinfo {author}
  {\bibfnamefont {K.}~\bibnamefont {McKenzie}},\ }\bibfield  {title} {\bibinfo
  {title} {High stability laser locking to an optical cavity using tilt
  locking},\ }\href {https://doi.org/10.1364/OL.427615} {\bibfield  {journal}
  {\bibinfo  {journal} {Opt. Lett.}\ }\textbf {\bibinfo {volume} {46}},\
  \bibinfo {pages} {3199} (\bibinfo {year} {2021})}\BibitemShut {NoStop}%
\bibitem [{\citenamefont {Al-Taiy}\ \emph {et~al.}(2014)\citenamefont
  {Al-Taiy}, \citenamefont {Wenzel}, \citenamefont {Preu{\ss}ler},
  \citenamefont {Klinger},\ and\ \citenamefont {Schneider}}]{37_AlTaiy2014}%
  \BibitemOpen
  \bibfield  {author} {\bibinfo {author} {\bibfnamefont {H.}~\bibnamefont
  {Al-Taiy}}, \bibinfo {author} {\bibfnamefont {N.}~\bibnamefont {Wenzel}},
  \bibinfo {author} {\bibfnamefont {S.}~\bibnamefont {Preu{\ss}ler}}, \bibinfo
  {author} {\bibfnamefont {J.}~\bibnamefont {Klinger}},\ and\ \bibinfo {author}
  {\bibfnamefont {T.}~\bibnamefont {Schneider}},\ }\bibfield  {title} {\bibinfo
  {title} {Ultra-narrow linewidth, stable and tunable laser source for optical
  communication systems and spectroscopy},\ }\href
  {https://doi.org/10.1364/OL.39.005826} {\bibfield  {journal} {\bibinfo
  {journal} {Opt. Lett.}\ }\textbf {\bibinfo {volume} {39}},\ \bibinfo {pages}
  {5826} (\bibinfo {year} {2014})}\BibitemShut {NoStop}%
\end{thebibliography}%


\providecommand{\noopsort}[1]{}\providecommand{\singleletter}[1]{#1}%
\begin{thebibliography}{2}%
\makeatletter
\providecommand \@ifxundefined [1]{%
 \@ifx{#1\undefined}
}%
\providecommand \@ifnum [1]{%
 \ifnum #1\expandafter \@firstoftwo
 \else \expandafter \@secondoftwo
 \fi
}%
\providecommand \@ifx [1]{%
 \ifx #1\expandafter \@firstoftwo
 \else \expandafter \@secondoftwo
 \fi
}%
\providecommand \natexlab [1]{#1}%
\providecommand \enquote  [1]{``#1''}%
\providecommand \bibnamefont  [1]{#1}%
\providecommand \bibfnamefont [1]{#1}%
\providecommand \citenamefont [1]{#1}%
\providecommand \href@noop [0]{\@secondoftwo}%
\providecommand \href [0]{\begingroup \@sanitize@url \@href}%
\providecommand \@href[1]{\@@startlink{#1}\@@href}%
\providecommand \@@href[1]{\endgroup#1\@@endlink}%
\providecommand \@sanitize@url [0]{\catcode `\\12\catcode `\$12\catcode
  `\&12\catcode `\#12\catcode `\^12\catcode `\_12\catcode `\%12\relax}%
\providecommand \@@startlink[1]{}%
\providecommand \@@endlink[0]{}%
\providecommand \url  [0]{\begingroup\@sanitize@url \@url }%
\providecommand \@url [1]{\endgroup\@href {#1}{\urlprefix }}%
\providecommand \urlprefix  [0]{URL }%
\providecommand \Eprint [0]{\href }%
\providecommand \doibase [0]{https://doi.org/}%
\providecommand \selectlanguage [0]{\@gobble}%
\providecommand \bibinfo  [0]{\@secondoftwo}%
\providecommand \bibfield  [0]{\@secondoftwo}%
\providecommand \translation [1]{[#1]}%
\providecommand \BibitemOpen [0]{}%
\providecommand \bibitemStop [0]{}%
\providecommand \bibitemNoStop [0]{.\EOS\space}%
\providecommand \EOS [0]{\spacefactor3000\relax}%
\providecommand \BibitemShut  [1]{\csname bibitem#1\endcsname}%
\let\auto@bib@innerbib\@empty
\bibitem [{\citenamefont {Siegman}(1986)}]{25_Siegman1986}%
  \BibitemOpen
  \bibfield  {author} {\bibinfo {author} {\bibfnamefont {A.~E.}\ \bibnamefont
  {Siegman}},\ }\href@noop {} {\emph {\bibinfo {title} {Lasers}}}\ (\bibinfo
  {publisher} {University Science Books},\ \bibinfo {year} {1986})\BibitemShut
  {NoStop}%
\bibitem [{\citenamefont {Slagmolen}\ \emph {et~al.}(2002)\citenamefont
  {Slagmolen}, \citenamefont {Shaddock}, \citenamefont {Gray},\ and\
  \citenamefont {McClelland}}]{38_Slagmolen2002}%
  \BibitemOpen
  \bibfield  {author} {\bibinfo {author} {\bibfnamefont {B.}~\bibnamefont
  {Slagmolen}}, \bibinfo {author} {\bibfnamefont {D.}~\bibnamefont {Shaddock}},
  \bibinfo {author} {\bibfnamefont {M.}~\bibnamefont {Gray}},\ and\ \bibinfo
  {author} {\bibfnamefont {D.}~\bibnamefont {McClelland}},\ }\bibfield  {title}
  {\bibinfo {title} {Frequency stability of spatial mode interference (tilt)
  locking},\ }\href {https://doi.org/10.1109/JQE.2002.804267} {\bibfield
  {journal} {\bibinfo  {journal} {IEEE Journal of Quantum Electronics}\
  }\textbf {\bibinfo {volume} {38}},\ \bibinfo {pages} {1521} (\bibinfo {year}
  {2002})}\BibitemShut {NoStop}%
\end{thebibliography}%

\end{document}


\preprint{APS/123-QED}

\title{Supplemental Material: Laser-cavity locking at the $10^{-7}$ instability scale utilizing beam elipticity}

\author{Fritz Diorico}
\author{Artem Zhutov}%
\author{Onur Hosten}%
 \email{onur.hosten@ist.ac.at}
\affiliation{%
Institute of Science and Technology Austria, Klosterneuburg, Austria}%


\maketitle

This supplemental material includes additional details on experimental implementation and characterization, derives the central formulas used in the main text, and discusses further applications. 

\tableofcontents

\subsection{The cavity setup and the alignment procedure} 

The cavity is made of one high reflectivity curved mirror (5-\SI{}{\centi\meter} radius of curvature) and two lower reflectivity ($\sim98.93$  for p-polarization at \SI{30}{\degree} angle of incidence) plane mirrors. The reflectivities are chosen to make the cavity linewidth larger than those of the lasers (\SI{500}{\kilo\hertz}) for simplicity. The mirrors are attached with an epoxy resin on an aluminum block with three drilled bores that form a nearly equilateral triangle. The FC and CLP (Fig. 3a) are mounted on a cage assembly, which is situated on a 5-axis tip-tilt-translation stage. This assembly, together with the axial translation capability of the FC lens allows us to mode match to the cavity. The cavity, together with the remaining optics are all mounted directly onto the optical table, and are enclosed with a metal box for temperature and air fluctuation stability. The aluminum cavity spacer is temperature stabilized to within \SI{1}{\milli\kelvin} to control the cavity length.

To experimentally identify the cavity mode reference profile (Fig. 3b), light is back coupled to the ‘00’ mode from one of the mirrors and the mode profile emitted out of the cavity (along the path to the QPDs) is characterized with a CMOS camera. Incident beam profile is adjusted by tuning the location of the optical elements to match the reference mode upon reflection from the cavity. Once aligned, all other modes but the ‘00’ and the ‘02’-‘20’ are suppressed, with an equal transmission level for the latter two.

Following the input light alignment, the QPD location is aligned. For centering the beam on the QPD, the DL is translated while monitoring the left-right and up-down signals from the QPD. With the beam centered, a baseline offset on the error signal is attributable to an axial displacement of the QPD location from the location where the beam assumes a circular profile. The QPD location can in principle be adjusted to eliminate this offset. However, the natural astigmatism of the cavity also contributes to the baseline offset. This also gives rise to a shift of the transmission peak from the error signal zero-crossing. Hence the axial QPD location can be adjusted until the transmission peak and the error signal zero-crossing matches. In absence of this matching, a cross-coupling of intensity noise to frequency noise emerges.

\subsection{The cylindrical lens pair}

The two cylindrical lenses have identical focal lengths, with one oriented horizontally and the other vertically. The focal length is chosen such that if the two lenses were co-located (as if they formed one spherical lens), the incident beam would be perfectly mode matched to the ‘00’ cavity mode. As each member of the pair is translated symmetrically in opposite directions from this reference configuration, the ‘02’ and ‘20’ modes are populated with equal amplitudes as observed from the cavity transmission spectrum (Fig. 3c). Physically, this action translates the beam waists for the horizontal and vertical planes to before and after the cavity waist location (Fig. 3b). 

One can think of the beam shape evolution along the beam path as an evolving interference between the ‘00’, ‘02’ and ‘20’ components as these components acquire different amounts of Gouy phases \cite{25_Siegman1986}. In this picture, at the location of the QPD, the Gouy phase shifts are such that we obtain an equal superposition of ‘02’ and ‘20’ with a $\pi$-radian relative phase, forming the ‘+’ component depicted in Fig. 1b. Furthermore, the relative phase of the ‘00’ and the ‘+’ component becomes $\pi/2$ radians – when off-resonant – as described in the main text. In practice, we do not calculate Gouy phase shifts, but simply propagate Gaussian beams using the ABCD matrices \cite{25_Siegman1986} of the optical setup. With this calculation, as depicted in Fig. 3b, we identify the variation of spot sizes for the horizontal and vertical planes, the location where the beam becomes circular, and the beam size at this location. Note that all important properties of the beam are determined by the CLP. The purpose of the detection lens is to merge the spot sizes in the horizontal and vertical planes (Fig. 3a\&b), and this happens regardless of its exact location or its focal length for a correctly configured CLP.

The detected error signal level is proportional to the product of the amplitudes in the ‘00’ and the ‘+’ components, since it arises as an interference effect between these two components on the QPD. For an input power normalized to unity, the magnitude of these amplitudes are given by $|a_{00}|=(1+(\frac{d}{4z_r})^2)^{-1/2}$  and  $|a_{+}|=\frac{d}{4z_r}(1+(\frac{d}{4z_r})^2)^{-3/2}$. Here $d$ is the CLP separation, and $z_R$ (\SI{2.2}{\centi\meter}) is the Rayleigh length associated with the cavity modes. Note that these results do not directly depend on the CLP focal lengths, since these are experimentally linked to $z_R$. The normalized amplitude product determining the error signal level is maximized to  $(|a_{00}||a_+|))_{max}=0.325$ at the optimum separation $d_{opt} = \frac{4}{\sqrt{3}}z_r$ (\SI{5.1}{\centi\meter}). We operate at $d = \SI{4}{\centi\meter}$, practically maximizing the signal for a given input power.

\subsection{The electronic backend}

The QPD and the feedback circuits are home built. All quadrants of the QPD are independently amplified with a \SI{380}{\kilo\ohm} transimpedance gain, and arithmetic operations are then carried out with operational amplifier circuits to simultaneously access the diagonal, left-right, up-down and sum signals. These respectively give us the cavity lock error signal (Fig. 1c), the beam centering signals and the total light intensity reflecting from the cavity. 

The signal from the diagonal channel goes to an analog feedback circuit and is fed to the laser current to keep the lasers on resonance with the cavity when locked. The feedback loop contains two integrators ($1/f^2$ response) from DC to \SI{20}{\kilo\hertz}, and a single integrator ($1/f$ response) from \SI{20}{\kilo\hertz} to the lock bandwidth of \SI{100}{\kilo\hertz}, providing a lock tightness that does not pose a dominant limitation to stability (crosses in Figure 4c). All input and output signals to the feedback circuits are differential in order to eliminate ground loops, which is crucial to performance. 

We note that the QPD together with the implemented amplification circuit operates shot-noise limited for frequencies larger than \SI{300}{\hertz}. Below this frequency amplifier noise starts taking over. However the observed locking stability for timescales larger than \SI{10}{\milli\second} is not related either to shot noise or to amplifier noise, and this is typical for other experiments operating in such timescales.

\subsection{System performance and comparison}

Assuming equal performances for the two copies of the laser-cavity locking setup, the actual instability for one of the copies is given by $1/ \sqrt{2}$  of the measured laser beatnote instability since the independent noises add in quadrature. This correction is already taken into account in Figure 4, and is a standard procedure. This in general provides a conservative estimate of achievable instability, since if one of the two setups were to perform better, extracting the performance of that setup would require dividing with an even larger number. In our case, the performances of the two copies are roughly equal, since the characterized beam shape noise which limits the performance is roughly the same for both copies.

For comparison with previous state-of-the-art, we extract instability data from the works cited in Figure 4 using WebPlotDigitizer (https://automeris.io/WebPlotDigitizer), and scale the data to represent fractions relative to cavity linewidths utilized in each work. In choosing specific publications for comparison, we payed particular attention to works involving more than one similar setup to obtain a real measured stability through a laser beatnote. Note that, one can try to infer a locking stability from a single copy of the system by recording the in-loop error signal and utilizing an appropriate conversion factor. This method is not necessarily reliable since a feedback loop suppresses the error signal irrespective of its physical origin. For example, the in-loop error signal in our work continues to average into the $10^{-8}$ instability scale, but the real stability limitation is brought by the deviation of the error signal from the actual frequency value due to beam shape noise. Among the comparison data, the ‘Slagmolen’ and ‘Salomon’ curves, together with the results presented in this work are obtained by locking two lasers to the same cavity, whereas the ‘Matei’ curve is obtained by comparing two lasers locked to separate cavities – and is limited by thermal the noise of the utilized cavities. 

An advantage of the locking method investigated in this work (shared with the PDH method) is its use of the reflected beam. Unlike in the ‘transmission modulation’ method (Fig. 4a), in this mode of operation, lock bandwidth is not directly limited by the cavity linewidth. This is important in the context of narrow linewidth high finesse cavities, when one tries to narrow the linewidth of a laser in addition to stabilizing its central frequency.

As an additional note, the fundamental limits of many frequency locking techniques due to optical shot noise have been shown to be the same to within a numerical factor of order unity \cite{38_Slagmolen2002}, and our method is no exception. While the ultimate limit is still given by shot noise, our method allows to more easily go towards this limit.

\subsection{Noise characterization}

The estimated limitations to stability presented in Figure 4c were carried out ex situ. The beam shape noise was measured by physically blocking the cavity mode, so that no light builds up inside the cavity. In this configuration, the input mirror promptly directs the incident light to the QPD and allows one to characterize the beam ellipticity fluctuations in the incident beam. No further modeling was done to take into account any shape noise filtering effects that the cavity might perform while the mode is not blocked. The QPD dark noise was measured by blocking the incident light and monitoring the error signal. The intensity drift noise was measured by simultaneously monitoring the laser beatnote frequency and the sum channels of the QPDs while the lasers were locked to the cavity. Except for the intensity noise, the on-resonance error signal slope was used to convert the measurements to frequency. For the intensity drift noise, the measured laser intensity dependence of the beatnote in locked operation was used. We note that there is no active intensity stabilization in this work.

It was verified that the stability limitation indeed comes from beam ellipticity fluctuations and not indirectly from beam pointing fluctuations combined with non-idealities in the QPD detection system. By translating the QPD position perpendicular to the beam axis, a cross-talk isolation between the beam position and beam ellipticity signals in excess of a factor of 200 was observed. On the other hand, the voltage noise in the beam position signals during operation are only one order of magnitude larger than that in the ellipticity signals, ruling out the beam pointing cross-talk as the dominating noise source.

A narrower laser linewidth would in principle improve stability in the short time scales, assuming there are no other dominating sources of detection noise in the system. In our case, artificially broadening the laser linewidth from \SI{500}{\kilo\hertz}  to \SI{1}{\mega\hertz} (by injecting current noise) did not deteriorate the performance, suggesting that the linewidth is not the limitation. We note that there is no laser linewidth narrowing induced in our parameter regime, since the lock bandwidth (\SI{100}{\kilo\hertz}) is smaller than the initial laser linewidth. 

\subsection{Advantages of post-selection}

In addition to the already-discussed suppression of sensitivity to intensity noise and incident beam shape fluctuations, the post-selection reduces sensitivity to beam pointing fluctuations. Although these fluctuations affect the signal in the diagonal QPD channel only to second order, and do not appear to cause a limitation in the current work, they are nevertheless further reduced. Another non-ideality that the post-selection suppresses is an adverse effect of the astigmatism of the triangular cavity modes. The astigmatism causes the zero-crossings of the error signal to become offset from the transmission intensity maxima. This results in reduced performance through intensity-to-frequency noise conversion. One can mitigate this effect by fine tuning alignment as discussed in the alignment procedures, nonetheless, the post-selection significantly suppresses these problems, rendering alignments easier.

Qualitatively, the improvements brought by the post-selection can be thought of as originating from a differential measurement performed on two beams with orthogonal polarizations – one that goes into the cavity and one that is promptly reflected from the cavity. Through post-selection, this differential measurement is manifested as a coherent process which improves the slope of the error signal (making it more sensitive to smaller frequency shifts) for a fixed amount of light impinging on the QPD. The post-selection process improves performance as long as the resulting intensity on the QPD does not fall below an equivalent electronic noise floor. 

Alongside all improvements, the post-selection is also observed to introduce a small sensitivity of frequency lock point to cavity length changes. Preliminarily characterizations suggest that this increased sensitivity is likely due to polarization changes in the light reflecting from the cavity as the light frequency changes – effectively changing the post-selection angle. Nevertheless cavity length is a parameter that is necessarily well controlled in frequency stabilization applications, and it does not pose a dominant limitation in our implementation.  

\subsection{Post selected beam evolution and the weak value of the reflection coefficient} 

Here we outline the derivation of the effective reflection coefficient and related formulas used in the main text. First, we model the cavity reflection without the polarization degree of freedom in the quantum mechanical language used for describing weak-values, then we take into account the polarization.

In the first case, there are two degrees of freedom in the problem: the ‘mode shape’ and the ‘channel’. The states spanned by the first are the various TEM modes, and the states spanned by the second are the labels ‘incident’ ($\ket{\mathcal{I}}$), ‘reflected’ ($\ket{\mathcal{R}}$) and ‘transmitted’ ($\ket{\mathcal{T}}$). We take the initial state as a direct product state for the two degrees of freedom: $\ket{\Psi}=\ket{\Phi} \otimes \ket{\mathcal{I}}$. Here, the state ‘$\Phi$' stands for an arbitrary input mode shape. To phenomenologically describe reflection from the cavity near the ‘00’ resonance, we define a unitary operator $\hat{U}=\dyad{0}{0}\otimes \hat{U}_C + (\mathbb{I} -\dyad{0}{0}) \otimes (\dyad{\mathcal{I}}{\mathcal{R}}+\dyad{\mathcal{R}}{\mathcal{I}})$  which evolves the state of the ‘channel’ with another unitary operator $\hat{U}_C$  if the ‘mode shape’ is in state  $\ket{0}$, and evolves the incident state  $\ket{\mathcal{I}}$ directly into the reflected state  $\ket{\mathcal{R}}$ otherwise. Here $\mathbb{I}$ is the identity operator for the ‘mode shape’ degree of freedom, and $\dyad{0}{0}$  is the projection operator for the ‘00’ mode. If we are only interested in the reflected part of the beam, an effective evolution operator $\hat{R}$  that acts only on the input ‘mode shape’ state  $\ket{\Phi}$ can be defined using a projection onto the reflected part after the state evolution: $\hat{R}\ket{\Phi}=\bra{\mathcal{R}}\hat{U}\ket{\Phi}\ket{\mathcal{I}}$. This procedure yields the effective evolution operator  $\hat{R}=r\dyad{0}{0} +(\mathbb{I}-\dyad{0}{0})$ with $r=\bra{\mathcal{R}}\hat{U}_C\ket{\mathcal{I}}$ being the amplitude reflection coefficient – an experimentally determined quantity.

Now, inclusion of the polarization degree of freedom introduces a more general unitary operator that evolves the initial state $\ket{\Psi}=\ket{\psi_1}\otimes\ket{\Phi}\otimes \ket{\mathcal{I}}$ . Here $\ket{\psi_1}$  is the initial state of the polarization degree of freedom. Using the above procedure, an effective evolution operator $\hat{R}=\hat{R}_H+\hat{R}_V$  acting only on the initial ‘mode shape’ and ‘polarization’ state $\ket{\psi_1}\otimes\ket{\Phi}$  can be deduced, yielding $\hat{R}_H=\dyad{H}{H}\otimes\left(r_H \dyad{0}{0} +(\mathbb{I}-\dyad{0}{0}) \right)$ and  $\hat{R}_V=\dyad{V}{V}\otimes e^{i\phi}\left(r_V \dyad{0}{0} +(\mathbb{I}-\dyad{0}{0})\right)$. In these expressions, $r_H$ and $r_V$ are the amplitude reflection coefficients for the horizontal $\ket{H}$  and vertical $\ket{V}$  polarization states – corresponding to the principal axes of the cavity. Here, a potential relative birefringent phase shift $\phi$ is explicitly included for the prompt reflection from an angled cavity input mirror. A further projection onto a polarization state $\ket{\psi_2'}$ using the definition $\hat{R}_{post}\ket{\Phi}=\bra{\psi_2'}\hat{R}\ket{\psi_1}\ket{\Phi}$, yields the final post-selected evolution operator   $\hat{R}_{post} =\braket{\psi_2}{\psi_1}[r_w\dyad{0}{0}+(\mathbb{I}-\dyad{0}{0})]$ which acts purely on the input mode state $\ket{\Phi}$. Here,  $r_w =\bra{\psi_2}\hat{r}\ket{\psi_1}/\braket{\psi_2}{\psi_1}$ is the weak-value of the reflection operator  $\hat{r}=r_H \dyad{H}{H} +r_V \dyad{V}{V}$, and the overlap coefficient $\braket{\psi_2}{\psi_1}$  in $\hat{R}_{post}$  further signifies the lossy nature of the post-selection operation. Note that for clarity, we have absorbed the effect of the birefringent phase shift into the definition of the post-selected state: $\ket{\psi_2}=(\dyad{H}{H}+\dyad{V}{V} e^{-i\phi})\ket{\psi_2 '}$. In the main text, the small birefringent phase is ignored, but the residual asymmetry of the curves in Fig. 2b can be attributed to it. 

Now we will evaluate  $r_w$ for our experimental configuration. With the use of linear polarizers, the pre- and post-selected states can be written as $\ket{\psi_1}=\cos{\theta_1}\ket{H}+\sin{\theta_1}\ket{V}$  and  $\ket{\psi_2}=\cos{\theta_2}\ket{H}+e^{-i\phi}\sin{\theta_2}\ket{V}$. This results into  $r_w = r_H (\frac{1}{2}+B)+r_V (\frac{1}{2}-B)$ for the weak-value, and  $\braket{\psi_2}{\psi_1}=\cos(\theta_1+\theta_2)/\left(B+\frac{1}{2}+e^{-i\phi}(B-\frac{1}{2})\right)$ for the overlap coefficient. Here we defined the complex number  $B=(1+e^{i\phi}\tan\theta_1\tan\theta_2)^{-1}-\frac{1}{2}$. For the present work the cavity is polarization resolving, and we can take the vertical polarization component to be far off-resonance, and use $r_H = 1 - \gamma/(1-i\frac{\delta}{\kappa/2})$  and $r_V =1$  with  $\gamma$, $\delta$  and  $\kappa$ as defined in the main text. This results into the weak value $r_w=1-\gamma(B+\frac{1}{2})/(1-i\frac{\delta}{\kappa/2})$. Lastly, we will make the small-birefringent-phase approximation $\phi\ll 1$  and expand in a power series in  $\phi$ to reach the formulas used in the main text. We obtain 
\begin{equation}
\label{eq:2}
r_w = 1 -\gamma'\frac{1}{1-i\frac{\delta}{\kappa/2}}, \tag{$S1$}
\end{equation}
\begin{equation}
\label{eq:3}
\braket{\psi_2}{\psi_1}=\frac{e^{-i\phi(A-1)}}{2A-1}\left(1 + (A^2-A)\frac{\phi^2}{2} \right)\cos(\theta_1+\theta_2)\tag{$S2$}
\end{equation}
where  $\gamma'=A\gamma e^{i\phi(A-1)}$ and  $A=\frac{1}{2}+B|_{\phi=0} = \frac{1}{2}\left(1 +\frac{cos(\theta_1+\theta_2)}{cos(\theta_1-\theta_2)} \right) $. The successful post-selection probability can be written as $|\braket{\psi_2}{\psi_1}|^2 \approx \frac{1}{(2A-1)^2} +\frac{(A^2-A)}{(2A-1)^2}\phi^2$, where we assumed $\cos(\theta_1+\theta_2)^2\sim1$ for the relevant post-selection angles. The expressions used in the main text further set $\phi=0$. 

\subsection{Temperature stability}

The observed locking instability of $5 \times 10^{-7}$ at \SI{10}{\second} averaging time is reached regardless of stringent temperature requirements. As the optical table temperature drifts in excess of \SI{100}{\milli\kelvin} over the course of an hour, this result is still achieved, and it is a matter of covering the setup with a cardboard box. Achieving the demonstrated instability levels of $3\times10^{-6}$ at the  \SI{1000}{\second} averaging time requires better temperature stabilities, obtained by enclosing the system in an additional metal box, reaching \SI{10}{\milli\kelvin} level drifts in the course of \SI{12}{\hour}. Without these measures, the locking instability at this averaging time climbs back up an order of magnitude larger than presented.

To serve as a guide for future improvements in long-term stability, we monitored the temperature near many elements on the optical table via thermistors, and further mounted resistive heaters at nearby locations to slowly alter temperatures in order to identify the most sensitive elements. We will summarize these below. We note that with these elements we also tried stabilizing temperatures locally on the table: Although we could obtain \SI{1}{\milli\kelvin} level temperature stabilities locally at the point of measurement, we noticed temperatures were still fluctuating at \SI{10}{\milli\kelvin} levels a few cm down from the point of measurement, inferring that a next version of the setup would significantly benefit from mounting the components on a compact platform that is thermally isolated from the optical table. This will enable an active temperature stabilization at levels much better than  \SI{10}{\milli\kelvin}, potentially improving the long term stability by an order of magnitude. 

The parts that are most sensitive to temperature variations are the output optical fibers, the stages used for beam pointing alignment, and the cavity itself. 

\emph{Fibers}: The polarizing fibers (coiled in a \SI{2}{\centi\meter} radius loop), which we obtained the best results with, still showed some residual fluctuations in the beam shape as a function of the temperature, which we identified to originate from the drifts in the polarization of light before exiting the fiber due to thermally induced stress. This effect, which cannot be eliminated by simply sending the beam through a polarizer at the output, showed $5 \times10^{-5}$ peak-to-peak level periodic beam aspect ratio changes at as the temperature is continuously swept.

\emph{Stages used for beam pointing alignment}: Thermally induced mechanical beam pointing changes are observed to turn into small beam ellipticity changes as the beam propagates through the setup, affecting the error signal. This is estimated to be at similar levels to the fiber-caused limitations in the long-term.

\emph{Cavity}: For the post-selected case, we observed the beatnote between the two lasers to depend on the exact resonance frequency of the cavity despite locking the lasers to two degenerate modes. This appears to be caused by frequency dependent polarization rotations upon reflection from the cavity, affecting the post-selection angle. This effect did not cause a dominant limitation in the final version given that we controlled the aluminum cavity spacer temperature easily at \SI{1}{\milli\kelvin} level. Note that in an absolute frequency stabilization application, this will never be a problem, since one needs to work with an ultra-stable length cavity to begin with.

\subsection{Applications to different types of cavities}

In this work, a ring cavity is utilized where the incident and reflected beams are traveling on separate paths. For different applications, different cavity designs might be required. For example, ultralow-vibration sensitivity typically requires working with two-mirror standing wave cavities. For this configuration nothing changes regarding the main method: Although the reflected beam travels on the same path as the incident beam, it can easily be separated, e.g., using a quarter-wave plate together with a polarizing beam splitter (as is usually done for the PDH method). Also note that the specific mode spectrum of a cavity does not play a role in the implementation; the only requirement for proper operation is for the ‘00’ mode not to be spectrally degenerate with the second-order modes, and thus the second order modes can have arbitrary generic resonance frequencies. The main method can also be used with waveguide-like cavities which require evanescent coupling, e.g., with dielectric whispering gallery mode cavities, or fiber rings. A well-established coupling strategy is the use of a prism, where the evanescent tails from the prism during total internal reflection couple light in-and-out of the cavity. The prism thus effectively interfaces waveguide-like structures into free space optics, and it appears like an input mirror of a free-space cavity for the purposes of applying the methods developed here.

For the case of a two-mirror cavity, the potential implementation of the post-selection procedure deserves a couple of remarks. Firstly, accomplishing the post-selection requires the cavity to be birefringent. This can be accomplished with either stress-induced birefringence, or with the use of crystalline mirror coatings. The second remark is that the polarization degree of freedom is already utilized to separate the beam paths in this cavity geometry. Nevertheless, it is possible to simultaneously accomplish a tunable post-selection by introducing polarization-dependent losses on the beam path in place of the mentioned quarter-wave plate, e.g., using Fresnel reflections from a glass plate.

\subsection{Compatibility with low-light-level measurements}

One of the advantages of the PDH method is its demonstrated compatibility with low-light level measurements. For example, it is possible to use light levels at or below one microwatt to prevent heating of cavity mirrors to avoid perturbing the length of an ultrastable reference cavity. Note that only the tone that resonantly enters the cavity counts towards this power, and the promptly reflected tones can be made sufficiently large to overcome electronic detection noise. Also note that this property is not shared with the transmission modulation method, where the sidebands do need to go through the cavity. The Squash locking method demonstrated in this work is fully compatible with low light levels. Here, we assess the current limitations on how low the input power to the cavity could be made, both with and without post-selection. Then we discuss how these could be further improved. As a remark, we point out that in the current demonstration, only $\sim\SI{1}{\micro\watt}$ is actually hitting the QPD (after the post-selection), and this power is dominated by the ‘+’ reference mode contribution (Fig. 2b, purple and dashed curves).

In the setup, out of the $\SI{400}{\micro\watt}$  laser power incident on the system, \SI{360}{\micro\watt} is in the ‘00’ mode and \SI{40}{\micro\watt} is in the ‘+’ mode. Of the total power in the ‘00’ mode, only \SI{180}{\micro\watt} is in the Horizontal polarization which resonantly couples to the cavity; everything else reflects promptly from the input mirror. This resonantly coupled ‘effective incident power’ would be the cause of mirror heating related stability degradation in a high finesse cavity. The question is: how low can this power be?

For the post-selected case reaching $5\times10^{-7}$ instability, we see that the QPD noise floor (1/f noise and DC drift) limitation to stability is a factor of $7$ below the achieved stability (red diamonds in Fig. 4c vs. red circles in Fig. 4b). Thus the incident power can be lowered by a factor of 7 before we electronic detection noise floor takes over, i.e., for this particular implementation, \SI{25}{\micro\watt} of effective incident power would be the lower limit. For the non-post-selected case achieving $2\times10^{-6}$ instability, note that the utilized total power was actually reduced from \SI{400}{\micro\watt} to \SI{50}{\micro\watt} in order not to saturate the QPD electronics, corresponding to a utilized effective incident power of \SI{22}{\micro\watt}. In analogy to the post-selected case, one finds that the lowest effective incident power before becoming detection noise limited would be \SI{4}{\micro\watt}.

The required effective incident light levels can be further reduced if one increases the amount of power in the ‘+’ mode by increasing the ellipticity. This can be done by separating the CLP further. Based on the information in the methods section ‘Cylindrical lens pair’, a factor of 2 additional power level reduction can be achieved without sacrificing sensitivity with the current design. One can easily achieve more than an order of magnitude reduction in effective incident powers if one inverts the power balance between the ‘+’ and ‘00’ modes by means of a phase plate mode converter to prepare a beam having mostly a ‘+’ component alongside a small ‘00’ contribution (but still coupling to the ‘00’ mode of the cavity). Alternatively one can reduce the relative amount of the ‘+’ mode contribution by making the beam less elliptical, and then lock to one of the second order modes (note that these resonances still provide a good error signal; Fig. 3c), in effect the using the ‘00’ mode as the phase reference. For the PDH method these two modalities correspond to having more power in the sidebands than that in the carrier, and locking to one of the sidebands respectively. 

Improving the transimpedance amplifier electronics by a factor of a few without sacrificing bandwidth could be possible. This would further reduce required power levels. Last but not least, it could be possible to eliminate the apparent need for larger incident powers in the post-selected case by developing a polarization dependent focusing scheme which will eliminate the undesirable power reduction in the ‘+’ mode while achieving the effective impedance matched configuration.


\bibliography{supp}